\documentclass[review,5p,times,twocolumn]{elsarticle}

\usepackage{xspace}
\usepackage{ifthen}
\usepackage{alltt}
\usepackage{amsmath}    
\usepackage{amsfonts}
\usepackage{amssymb}
\usepackage{latexsym}
\usepackage{balance}
\usepackage{booktabs}
\usepackage{enumerate}
\usepackage[T1]{fontenc} 
\usepackage{graphicx}
\usepackage[scaled=0.85]{helvet}
\usepackage[utf8]{inputenc} 
\usepackage{lscape}
\usepackage{listings} 
\usepackage{mathtools}
\usepackage{multicol} 
\usepackage{needspace}
\usepackage{stmaryrd}
\usepackage{times}
\usepackage[normalem]{ulem} 
\usepackage{textcomp}
\usepackage[modulo]{lineno} 
\usepackage{paralist} 
\usepackage{lstautogobble}
\usepackage[multiple]{footmisc}
\usepackage{xcolor}
\usepackage[most]{tcolorbox}

\usepackage{float}
\usepackage{array}
\usepackage{pdfpages, caption, geometry}
\usepackage{subcaption}
\usepackage{pifont} 
\usepackage{multirow}

\usepackage{enumitem} 

\newcommand{\id}[1]{$-$Id: scgPaper.tex 32478 2010-04-29 09:11:32Z oscar $-$}

\newcommand{\code}[1]{\texttt{#1}}

\newcolumntype{L}[1]{>{\raggedright\arraybackslash}m{#1}} 

\newboolean{showcomments}
\setboolean{showcomments}{false} 
\ifthenelse{\boolean{showcomments}}
 {
 	\newcommand{\nbc}[3]{
 		{\colorbox{#3}{\bfseries\sffamily\scriptsize\textcolor{white}{#1}}}
		{\textcolor{#3}{\sf\small$\blacktriangleright$\textit{#2}$\blacktriangleleft$}}}
	
 }{
 	\newcommand{\nbc}[3]{}
 	
 }

\newboolean{showedits}
\setboolean{showedits}{false} 
\ifthenelse{\boolean{showedits}}
{
	\newcommand{\del}[1]{\textcolor{red}{\sout{#1}}} 
	\newcommand{\nbe}[3]{
		{\colorbox{#3}{\bfseries\sffamily\scriptsize\textcolor{white}{#1}}}
		{\textcolor{#3}{\sf\small$\blacktriangleright$\textit{#2}$\blacktriangleleft$}}}
}{
	\newcommand{\del}[1]{} 
	
	\newcommand{\nbe}[3]{}
}

\ifthenelse{\boolean{showedits}}
{
  \newtcolorbox{inserted}{%
       title=Inserted text:,
       colframe=blue,colback=blue!5!white,
       breakable,
       leftrule=0mm, 
       bottomrule=0mm,
       rightrule=0mm,
       toprule=0mm,
       arc=0mm, outer arc=0mm,
       oversize
  }
  \newtcolorbox{deleted}{%
       title=Deleted text:,
       colframe=red,colback=red!5!white,
       breakable,
       leftrule=0mm, 
       bottomrule=0mm,
       rightrule=0mm,
       toprule=0mm,
       arc=0mm, outer arc=0mm,
       oversize
  }
  \newtcolorbox{refactored}{%
       title=Rewritten text:,
       colframe=blue,colback=red!5!white,
       breakable,
       leftrule=0mm, 
       bottomrule=0mm,
       rightrule=0mm,
       toprule=0mm,
       arc=0mm, outer arc=0mm,
       oversize
  }
}{

}

\definecolor{source}{gray}{0.9}
\newcommand{\seclabel}[1]{\label{sec:#1}}
\newcommand{\secref}[1]{\autoref{sec:#1}}
\newcommand{\figlabel}[1]{\label{fig:#1}}
\newcommand{\figref}[1]{\autoref{fig:#1}}
\newcommand{\tablabel}[1]{\label{tab:#1}}
\newcommand{\tabref}[1]{\autoref{tab:#1}}

\newcommand{\lstref}[1]{\autoref{lst:#1}}

\newcommand{\eg}{\emph{e.g.,}\xspace}
\newcommand{\ie}{\emph{i.e.,}\xspace}
\newcommand{\etal}{\emph{et al.}\xspace}
\newcommand{\etc}{\emph{etc.}\xspace}

\usepackage[pdftex,colorlinks=true,pdfstartview=FitV,linkcolor=black,citecolor=black,urlcolor=black,bookmarks=false]{hyperref}

\journal{Journal of Systems and Software}

\newcommand{\rqI}{What types of information are present in class comments? To what extent do information types vary across programming languages?}
\newcommand{\rqII}{Can machine learning be used to automatically identify class comment types according to \emph{CCTM}?}
\newcommand{\repFolder}[1]{Folder ``\href{https://github.com/poojaruhal/RP-class-comment-classification}{RP/#1}'' in the Replication package}
\newcommand{\repFile}[1]{File ``\href{https://github.com/poojaruhal/RP-class-comment-classification}{RP/#1}'' in the Replication package}

\newcommand{\finding}[2]{\vspace{2mm}
\noindent\fbox{\parbox{0.98\linewidth}{\fontsize{9}{10}\selectfont
\textbf{Finding #1.} #2 }}\\}

\begin{document}

\begin{frontmatter}

\title{How to Identify Class Comment Types? \\A Multi-language Approach for Class Comment Classification}

\author[SCG]{Pooja Rani}
\ead{pooja.rani@inf.unibe.ch}
\author[Zahw]{Sebastiano Panichella}
\ead{panc@zhaw.ch}
\author[SCG]{Manuel Leuenberger}
\ead{manuel.Leuenberger@inf.unibe.ch}
\author[RCOST]{Andrea Di Sorbo}
\ead{disorbo@unisannio.it}
\author[SCG]{Oscar Nierstrasz}
\ead{oscar.nierstrasz@inf.unibe.ch}

\address[SCG]{Software Composition Group, University of Bern, Switzerland}
\address[Zahw]{Zurich University of Applied Science, Switzerland}
\address[RCOST]{Department of Engineering, University of Sannio, Italy}

\begin{abstract}
Most software maintenance and evolution tasks require developers to understand the source code of their software systems.  
Software developers usually inspect class comments to gain knowledge about program behavior, regardless of the programming language they are using.
Unfortunately,
(i) different programming languages present language-specific code commenting notations and guidelines; and
(ii) the source code of software projects often lacks comments that adequately describe the class behavior, which complicates program comprehension and evolution activities.

To handle these challenges, this paper investigates the different language-specific class commenting practices of three programming languages: Python, Java, and Smalltalk. 
In particular, we systematically analyze the similarities and differences of the information types found in class comments of projects developed in these languages. 

We propose an approach that leverages two techniques ---\,namely Natural Language Processing and Text Analysis\,--- to automatically identify \emph{class comment types}, \ie the specific types of semantic information found in class comments.
To the best of our knowledge, no previous work has provided a comprehensive taxonomy of class comment types for these three programming languages with the help of a common automated approach.
Our results confirm that our approach can classify frequent class comment information types with high accuracy for the Python, Java, and Smalltalk programming languages.
We believe this work can help in monitoring and assessing the quality and evolution of code comments in different programming languages, and thus support maintenance and evolution tasks.
\end{abstract}


\end{frontmatter}

\section{Introduction}
\seclabel{introduction}
Software maintenance and evolution tasks require developers to perform program comprehension activities~\cite{Fjel83a,Haid10a}.
To understand a software system, developers usually refer to the software documentation of the system~\cite{Bavo13b,Souz05a}. Previous studies have demonstrated that developers trust code comments more than other forms of documentation when they try to answer program comprehension questions~\cite{Maal14a,Wood81a,Souz05a}.  
In addition, recent work has also demonstrated that \emph{``code documentation''}  is the most used source of information for bug fixing, implementing features, communication, and even code review~\cite{Mull13a}. 
In particular, well-documented code simplifies software maintenance activities, but many programmers often overlook or delay code commenting tasks~\cite{Curi13a}.

Class comments play an important role in obtaining a high-level overview of the classes in object-oriented languages~\cite{Clin15a}. 
In particular, when applying code changes, developers using object-oriented programming languages can inspect class comments to achieve most or the majority of the high-level insights about the software system design, which is critical for program comprehension activities~\cite{Kham10a,Nurv03a,Stei13b}.    
Class comments contain various types of information related to the usage of a class or its implementation~\cite{Haou11a}, which can be useful for other developers performing program comprehension~\cite{Wood81a} and software maintenance tasks~\cite{Souz05a}.
Unfortunately, (i) different object-oriented programming languages adopt language-specific code commenting notations and guidelines~\cite{Faro15a}, (ii) they embed different kinds of information in the comments~\cite{Scow74a,Ying05a}; and (iii) software projects commonly lack comments to adequately describe the behavior of classes, which complicates program comprehension and evolution activities~\cite{More13c,Pani16a,Pani12a}.
The objective of our work is to examine developer class commenting practices (\eg comment content and style) across multiple languages, investigating the way in which developers write information in comments, and to establish an approach to identify that information in a language-independent manner.
To achieve this objective, we select three representative programming languages based on
\begin{inparaenum}[i)]
    \item whether the language is statically- or dynamically-typed,
    \item the level of detail embedded in its class comments,
    \item whether it supports live programming, and 
    \item the availability of a code comment taxonomy for the language (investigated in previous work).
\end{inparaenum}
The motivation behind each criterion is explained in the following paragraphs.

As mentioned, programming languages adopt different conventions to embed various types of information in class comments.
For example, in Java (a statically-typed language), a class comment provides a high-level outline of a class, \eg a summary of the class, what the class actually does, and other classes it cooperates with~\cite{Nurv03a}.
In Python (a dynamically-typed language), the class comment guidelines suggest adding low-level details about public methods, instance variables, and subclasses, in addition to the high-level summary of the class~\cite{Pyth20a}.\footnote{\url{https://www.python.org/dev/peps/pep-0257/}}
In Smalltalk (another dynamically-typed language), class comments are a primary source of code documentation, and thus they contain high-level design details as well as low-level implementation details of the class, \eg rationale behind the class, its instance variables, and important implementation points of its public methods.
By analyzing multiple languages that vary in the details of their class comments, we can provide a more general overview of class commenting practices than by focusing on a single language.

Programming languages offer various conventions and tools to express these different information types in class comments.
For example, Java supports Javadoc documentation comments in which developers use specific annotations, such as \emph{@param} and \emph{@author} to denote a given information type.
In Python, developers write class comments as docstrings containing similar annotations, such as \emph{param:} and \emph{args:} to denote various information types, and they use tools such as Pydoc and Sphinx to process these docstrings.
In contrast to Java and Python, class comments in Smalltalk neither use annotations nor the writing style of Javadoc or Pydoc, thus presenting a rather different perspective on commenting practices, and particular challenges for existing information identification approaches.
Additionally, Smalltalk is considered to be a pure object-oriented programming language and supports live programming since its inception, therefore, it can present interesting insights into code documentation in live programming environments.

We argue that the extent to which class commenting practices vary across different languages is an aspect only partially investigated in previous work.  
Given the multi-language nature of contemporary software systems, this investigation is critical to monitor and assess the quality and evolution of code comments in different programming languages, which is relevant to support maintenance and evolution tasks.
Therefore, we formulate the following research question:
\begin{quotation}
\noindent
\textbf{RQ$_1$} ``\emph{\rqI}''
\end{quotation}

We focus on addressing this question, as extracting class comment information types (or simply \emph{class comment types}) can help in providing custom details to both novice and expert developers, and can assist developers at different stages of development. 
Hence, we report the first empirical study investigating the different language-specific class commenting practices of three different languages: Python, Java, and Smalltalk. Specifically, we quantitatively and qualitatively analyze the class comments that characterize the information types typically found in class comments of these languages.
Thus, we present a taxonomy of class comment types based on the mapping of existing comment taxonomies, relevant for program comprehension activities in each of these three languages,  called \emph{CCTM} (\emph{\textbf{C}lass \textbf{C}omment \textbf{T}ype \textbf{M}odel}), mined from the actual commenting practices of developers.
Four authors analyzed the content of comments using card sorting and pair sorting~\cite{Guzz13a} to build and validate the comment taxonomy.

In the cases a code comment taxonomy was already available from previous works~\cite{Pasc17a,Zhan18a,Rani21b}, we used that taxonomy and refined it according to our research goals.

Our work provides important insights into the types of information found in the class comments of the investigated languages, highlighting their differences and commonalities.
In this context, we conjecture that these identified information types can be used to explore automated methods capable of classifying comments according to CCTM, which is a relevant step towards the automated assessment of code comments quality in different languages.
Thus, based on this consideration, we formulate a second research question:
\begin{quotation}
\noindent
\textbf{RQ$_2$} ``\emph{\rqII}"
\end{quotation}

To answer this research question, we propose an approach that leverages two techniques ---\,namely Natural Language Processing (NLP) and Text Analysis (TA)\,--- to automatically classify class comment types according to CCTM. Specifically, by analyzing comments of different types using NLP and TA we infer relevant features characterizing the class comment types.
These features are then used to train machine learning models enabling the automated classification of the comment types composing CCTM. 
 
Our results confirm that the use of these techniques allows class comments to be classified with high accuracy, for all investigated languages.
As our solution enables the presence or absence of different types of comment information needed for program comprehension to be determined automatically, we believe it can serve as a crucial component for tools to assess the quality and evolution of code comments in different programming languages.
Indeed, this information is needed to improve code comment quality and to facilitate subsequent maintenance and evolution tasks.

In summary, this paper makes the following contributions:
\begin{enumerate}
    \item an empirically validated taxonomy, called \emph{CCTM}, characterizing the  information types found in class comments written by developers in three different programming languages,
    	\item an automated approach (available for research purposes) able to classify class comments with high accuracy according to CCTM, and
    \item a publicly available dataset of manually dissected and categorized class comments in the replication package.\footnote{\fontsize{6}{6}\selectfont \url{https://github.com/poojaruhal/RP-class-comment-classification}}
\end{enumerate}


\section{Background}
\seclabel{background}
Code commenting practices vary across programming languages, depending on the language paradigm, the communities involved, the purpose of the language, and its usage in different domains \etc

\begin{figure*}[th]
    \centering
    \begin{minipage}{.47\linewidth}
        \centering
        \includegraphics[width=\linewidth,height=0.7\linewidth]{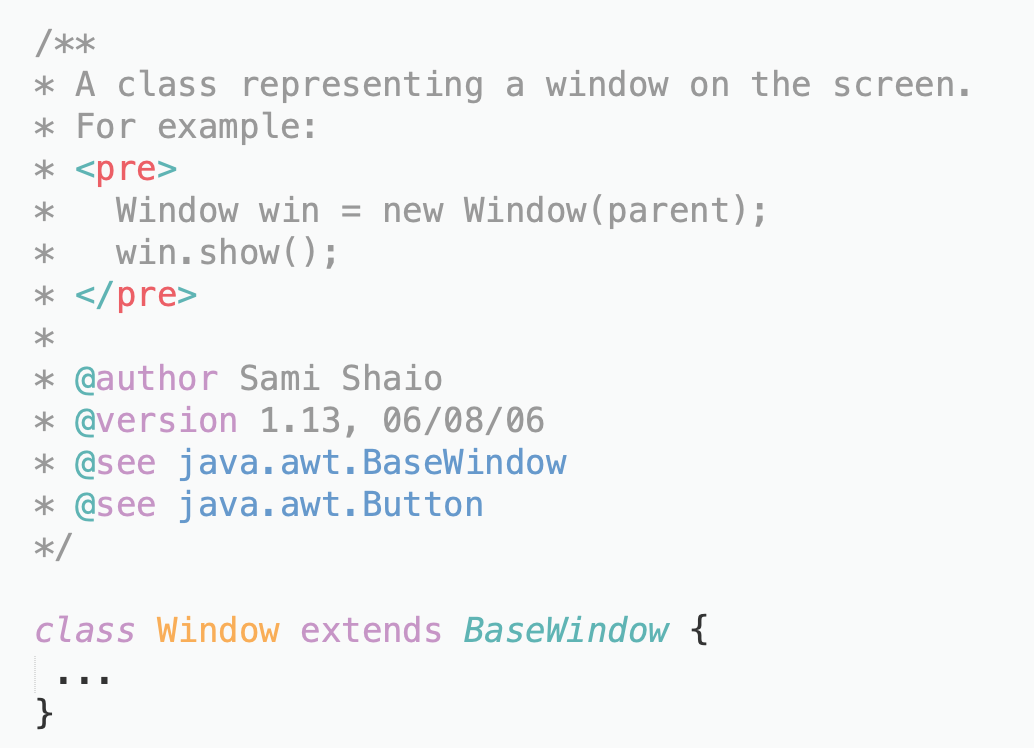}
        \caption{A class comment in Java}
        \figlabel{java-class-comment}
    \end{minipage}\hspace{4mm}
    \begin{minipage}{.47\linewidth}
        \centering
        \includegraphics[width=\linewidth,height=0.7\linewidth]{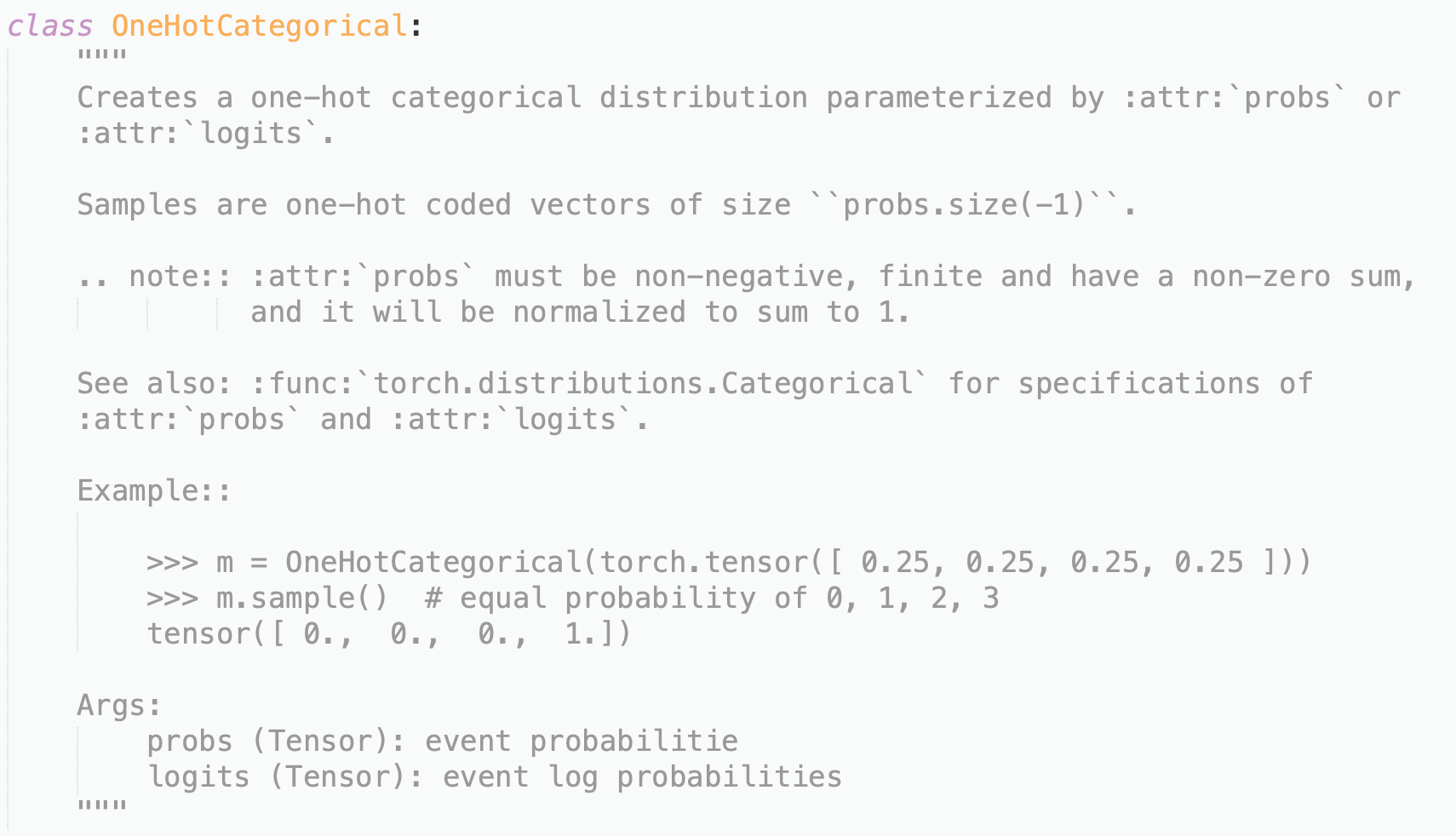}
        \caption{A class comment in Python}
        \figlabel{python-class-comment}
    \end{minipage}
\end{figure*}

While Java is a general-purpose, statically-typed object-oriented programming language with wide adoption in industry, Python, on the other hand, is dynamically-typed and supports object-oriented, functional, and procedural programming. 
We can observe differences in the notations used by Java and Python developers for commenting source code elements.
For instance, in Java, a class comment, as shown in \figref{java-class-comment}, is usually written above the class declaration using annotations (\eg @param, @version, \etc), whereas a class comment in Python, as seen in \figref{python-class-comment}, is typically written below the class declaration as ``docstrings''.\footnote{\url{https://sphinxcontrib-napoleon.readthedocs.io/en/latest/example_numpy.html}}
In Java, developers use dedicated Javadoc annotations, such as \emph{@author} and \emph{@see}, to denote a given information type.
Similarly, in Python, developers use similar annotations in docstrings, such as \emph{See also:}, \emph{Example:} and \emph{Args:}, and they use tools such as Pydoc and Sphinx to process them.

\begin{figure}[th]
    \centering
    \includegraphics[width=\linewidth]{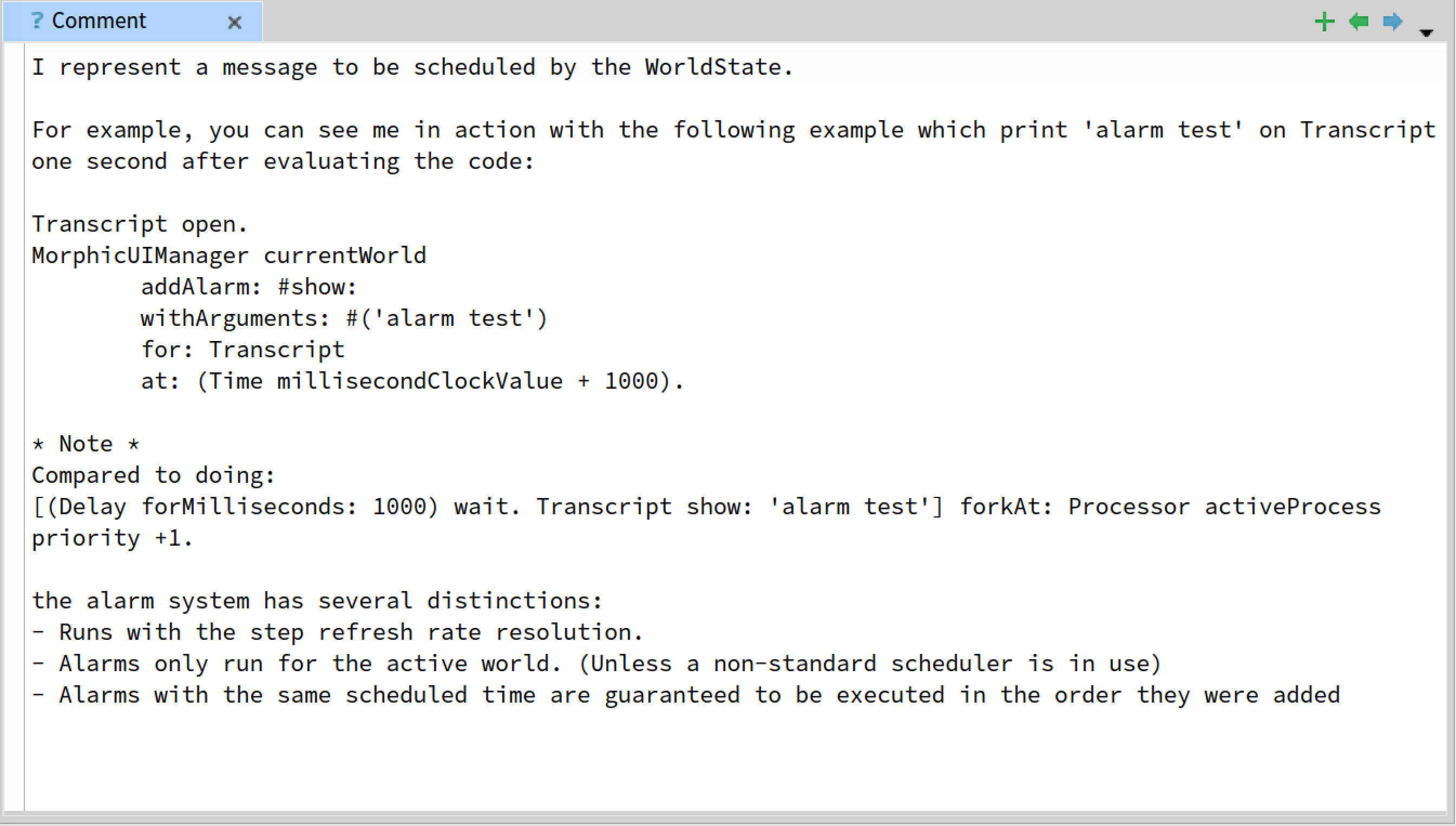}
    \caption{A class comment in Smalltalk}
    \figlabel{smalltalk-class-comment}
    \vspace{-2mm}
\end{figure}
Smalltalk is a pure, object-oriented, dynamically-typed, reflective programming language.
Pharo is an open-source and live development environment incorporating a Smalltalk dialect.
The Pharo ecosystem includes a significant number of projects used in research and industry~\cite{Pharo}.
A typical class comment in Smalltalk environment (Pharo) is a source of high-level design information about the class as well as low-level implementation details.
For example, the class comment of the class \emph{Morphic\-Alarm} in \figref{smalltalk-class-comment} documents
(i) the intent of the class (mentioned in the first line),
(ii) a code example to instantiate the class,
(iii) a note explaining the corresponding comparison, and
(iv) the features of the alarm system (in the last paragraph).
The class comment in Pharo appears in a separate pane instead of being woven into the source code of the class.
The pane contains a default class comment template, which follows a CRC (Class-Responsibility-Collaboration) model, but no other standard guidelines are offered for the structure and style of the comments. The comments follow a different, and informal writing style compared to Java, Python, and C/C++.
For instance, a class comment uses complete sentences, often written in the first-person form, and does not use any kind of annotations, such as \emph{@param} or \emph{@see} to mark the information type, as opposed to class comments in other languages~\cite{Nurv03a,Padi09a,Zhan18a}.
As a descendant of Smalltalk-80, Smalltalk has a long history of class comments being isolated from the source code~\cite{Gold83a}.
Class comments are the main source of documentation in Smalltalk.
Additionally, Smalltalk supports live programming for more than three decades, and therefore can present interesting insights into code documentation in a live programming environment.

Given the different commenting styles and the different types of information found in class comments from heterogeneous programming languages, the current study has the aim of (i) systematically examining the class commenting practices of different environments in details, and (ii) proposing an automated approach to identify the information contained in class comments.

\section{Study Design}
\seclabel{study-design}

\begin{figure*}[ht]
	\centering
	\includegraphics[width=\textwidth]{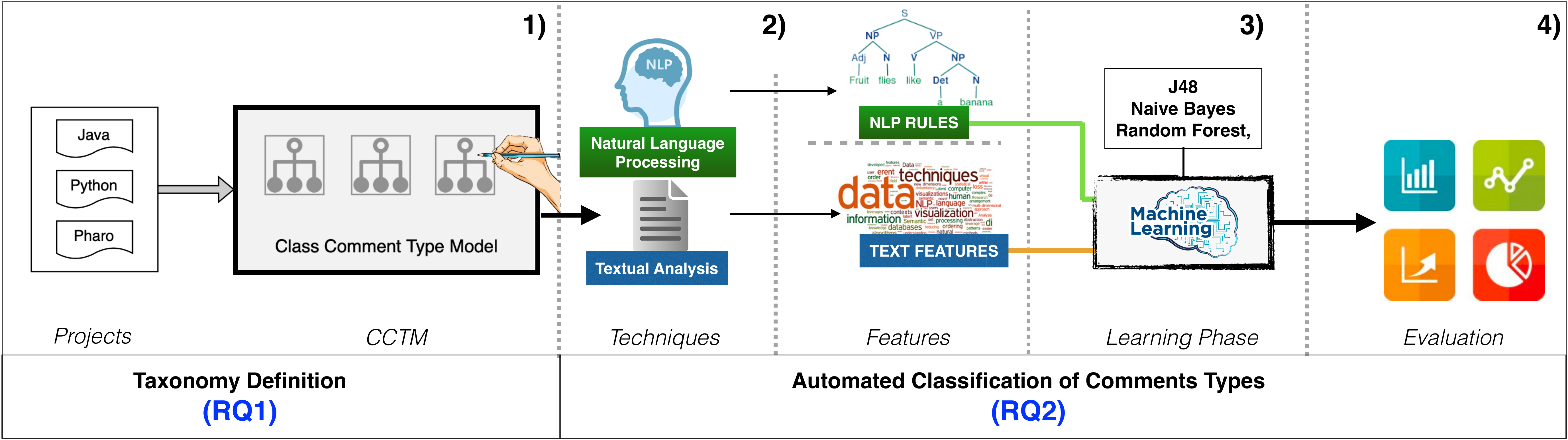}
	\vspace{-5mm}
    \caption{Overview Research Approach}
    \figlabel{approach}
    \vspace{-2.5mm}
\end{figure*}

The goal of our study is to understand the class commenting practices of developers across different object-oriented programming languages.
With the obtained knowledge, we aim to build a recommender system that can automatically identify the different types of information in class comments.
Such a system can provide custom details to both novice and expert developers, and assist them at various stages of development.  
\figref{approach} illustrates the research approach we followed to answer research questions RQ$_1$ and RQ$_2$.

\subsection{Data Collection}
\seclabel{data-collection}
For our investigation, we selected (i) Java and Python, two of the top  programming languages according to Google Trend popularity index\footnote{\url{https://pypl.github.io/PYPL.html}} and TIOBE index\footnote{\url{https://www.tiobe.com/tiobe-index/}}, and (ii) Smalltalk, as its class commenting practices emerged from Smalltalk-80~\cite{Pharo, Gold80b}.
Other criteria to select these languages are explained in the Introduction (\secref{introduction}) and Background section (\secref{background}).
Smalltalk is still widely used and it gained second place for \emph{most loved programming language} in the Stack Overflow survey of 2017.\footnote{\url{https://insights.stackoverflow.com/survey/2017/} verified on 4 Feb 2020}
For each language, we chose popular, open-source, and heterogeneous projects.
Such projects vary in terms of size, contributors, domains, ecosystems, and coding style guidelines (or comment guidelines).

As not all classes contain a class comment, we identified the classes having class comments for each project.
Afterwards, we extracted a statistically significant sample of class comments to conduct a manual analysis.
This analysis aimed at identifying the semantic information developers embed in class comments.
To determine the minimum size of the statistically significant sample for each language dataset, we set the confidence level to 95\% and the margin of error to 5\%~\cite{Trio06a}.


After determining the sample size, we selected the number of comments to consider from each project based on the proportion of the project's class comments of all comments (from all projects).
For instance, class comments from an Eclipse project in Java contribute to 29\% of total comments (comments from all Java projects) therefore we selected the same proportion of sample comments, \ie 29\% of Java sample size from the Eclipse project (110 class comments), as shown in \tabref{java_projects}.
To select representative sample comments from each project, we applied the stratified random sampling strategy and selected a proportional number of comments from each stratum.
The strata were defined based on the length (in terms of lines) of comments.
In particular, for each project, we first computed the quintiles based on the distribution of comments' length and treated the quintiles as strata.
For example, to choose 110 sample comments for Eclipse as shown in \tabref{java_projects}, we explored the distribution of comments lines and obtained quintiles as follows 1, 3, 4, 5, 7, and 1473.
Hence the five strata are 1-3, 4-4, 5-5, 6-7, and 8-1473.
Then from each stratum, we selected the proportional number of comments.

\textbf{Java:}
we selected six open-source projects analyzed in previous work~\cite{Pasc17a} to ease the comparison of our work with previous achievements.\footnote{\repFolder{Dataset/RQ1/Java}}
Modern complex projects are commonly developed using multiple programming languages.
For example, Apache Spark contains 72\% Scala classes, 9\% Java classes, and 19\% classes from other languages.\footnote{\url{https://github.com/apache/spark}}
In the context of our study, we considered the classes from the language under investigation, and discarded the classes from other programming languages.
For each class, we parsed the Java code and extracted the code comments preceding the class definition using the AST (Abstract Syntax Tree) based parser.
During extraction, we found instances of block comments (comments starting with /* symbol) in addition to Javadoc class comments (comments starting with /** symbol) before the class definition.
In such cases, the AST-based parser detects the immediately preceding comment (being it is a block comment or Javadoc comment) as a class comment and treats the other comment above it as a dangling comment.
To not miss any kinds of class comment, we adapted our parser to join both comments (the AST detected class comment and the dangling comment) as a whole class comment.
\begin{table}[ht]
    \centering
    \small
    \captionsetup{font=small}
    \caption{Overview of Java Projects}
\tablabel{java_projects}
\begin{tabular}{p{0.09\linewidth}p{0.15\linewidth}p{0.09\linewidth}p{0.16\linewidth}p{0.09\linewidth}p{0.16\linewidth}}
\hline\noalign{\smallskip}
\textbf{Project} &  \textbf{\% of Java classes} & \textbf{\#Java classes} & \textbf{\#Class comments} &\textbf{\% of dataset} & \textbf{\#Sample comments} \\
\noalign{\smallskip}\hline\noalign{\smallskip}
    Eclipse      &98\%      & 9\,128  & 6\,253  & 29\%  & 110  \\
    Spark        &9.3\%     & 1\,090  & 740     & 3.4\% &  13  \\
    Guava        &100\%     & 3\,119  & 2\,858  & 13\%  &  50  \\
    Guice        &99\%      & 552     & 466     & 2.1\% &  10  \\
    Hadoop       &92\%      & 11\,855 & 8\,846  & 41\%  &  155  \\
    Vaadin       &55\%      & 5\,867  & 2\,335  & 11\%  &  41  \\
    \noalign{\smallskip}\hline
\end{tabular}
\end{table}

We present the overview of the projects selected for Java in \tabref{java_projects} and raw files in the replication package.\footnote{\repFolder{Dataset/RQ1/Java}}
We established 376 class comments as the statistically significant sample size based on the total number of classes with comments.
For each project, the sample of class comments (the column \emph{Sample comments}) is measured based on the proportion of class comments in the total dataset of class comments (shown in the column \emph{``\% of dataset''} of \tabref{java_projects}).

The number of lines in class comments varies from 1 to 4605 in Java projects.
For each project, we established strata based on the identified quintiles from the project's distribution. 
From each stratum, we picked an equal number of comments using the random sampling approach without replacement.
For example, in the Eclipse project, we identified the five strata as 1-3, 4-4, 5-5, 6-7, and 8-1473.
We picked 25 comments from each stratum summing to the required 110 sample comments.
We followed the same approach for Python and Smalltalk to select the representative sample comments.

\textbf{Python:}
 We selected seven open-source projects, analyzed also in the previous work~\cite{Zhan18a}, to ease the comparison of our work with previous achievements.
To extract class comments from Python classes, we implemented a Python AST based parser and extracted the comments preceding the class definition.
\begin{table}[ht]
    \centering
    \small
    \captionsetup{font=small}
    \caption{Overview of Python Projects}
    \tablabel{python_projects}
    \begin{tabular}{cp{0.11\linewidth}p{0.16\linewidth}p{0.11\linewidth}p{0.16\linewidth}}
        \hline\noalign{\smallskip}
        \textbf{Project} &\textbf{\#Python classes} & \textbf{\#Class comments} &\textbf{\% of dataset} & \textbf{\#Sample comments} \\
        \noalign{\smallskip}\hline\noalign{\smallskip}
        Requests & 79     & 43     & 1.1\% & 4  \\
        Pandas   & 1753   & 377    & 9.9\% & 35  \\
        Mailpile & 521    & 283    & 7.5\% & 26  \\
        IPython  & 509    & 240    & 6.3\% & 22  \\
        Djnago   & 8\,750 & 1\,164 & 30\%  & 107  \\
        Pipenev  & 1\,866 & 1\,163 & 30\%  & 107  \\
        Pytorch  & 2\,699 & 520    & 13\%  & 48  \\
        \noalign{\smallskip}\hline
    \end{tabular}
\end{table}

The metadata related to the selected projects for Python are reported in \tabref{python_projects}, while the class comments are found in our replication package.\footnote{\repFolder{Dataset/RQ1/Python}}
We measured 349 sample comments to be the statistically significant sample size based on total classes with comments.

\textbf{Smalltalk:}
We selected seven open-source projects.
As with Java and Python, we selected the same projects investigated in previous research~\cite{Rani21b}.
\begin{table}[ht]
    \centering
    \small
   \captionsetup{font=small}
   \caption{Overview of Smalltalk Projects}
    \tablabel{pharo_projects}
    \begin{tabular}{cp{0.11\linewidth}p{0.16\linewidth}p{0.11\linewidth}p{0.16\linewidth}}
        \hline\noalign{\smallskip}
        \textbf{Projects} &\textbf{Total classes} & \textbf{\#Classes comments} &\textbf{\% of dataset} & \textbf{\#Sample comments} \\
        \noalign{\smallskip}\hline\noalign{\smallskip}
        GToolkit    & 4\,191& 1\,315& 43\% & 148  \\
        Seaside     & 841   & 411   & 14\% &  46  \\
        Roassal     & 830   & 493   & 16\% &  56  \\
        Moose       & 1\,283& 316   & 10\% &  36  \\
        PolyMath    & 300   & 155   & 5\%  &  17  \\
        PetitParser & 191   & 99    & 3\%  &  11  \\
        Pillar      & 420   & 237   & 8\%  &  27  \\
        \noalign{\smallskip}\hline
    \end{tabular}
\end{table}
\tabref{pharo_projects} shows the details of each project with the total numbers of classes, as well as classes with comments,\footnote{\repFolder{Dataset/RQ1/Pharo}} their proportion in the total comments, and the numbers of sample comments selected for our manual analysis.
We extracted the stable version of each project compatible with Pharo version 7 except for GToolkit, due to the lack of backward compatibility. 
We, therefore, used Pharo 8 for GToolkit.

\subsection{Analysis Method}
\seclabel{analysis-method}

\emph{RQ$_1$: \rqI}

The purpose of RQ$_1$ is to quantitatively and qualitatively  investigate the language-specific class commenting practices characterizing programs written in Python, Java, and Smalltalk.
\begin{figure}[th]
    \centering
    \includegraphics[width=\linewidth]{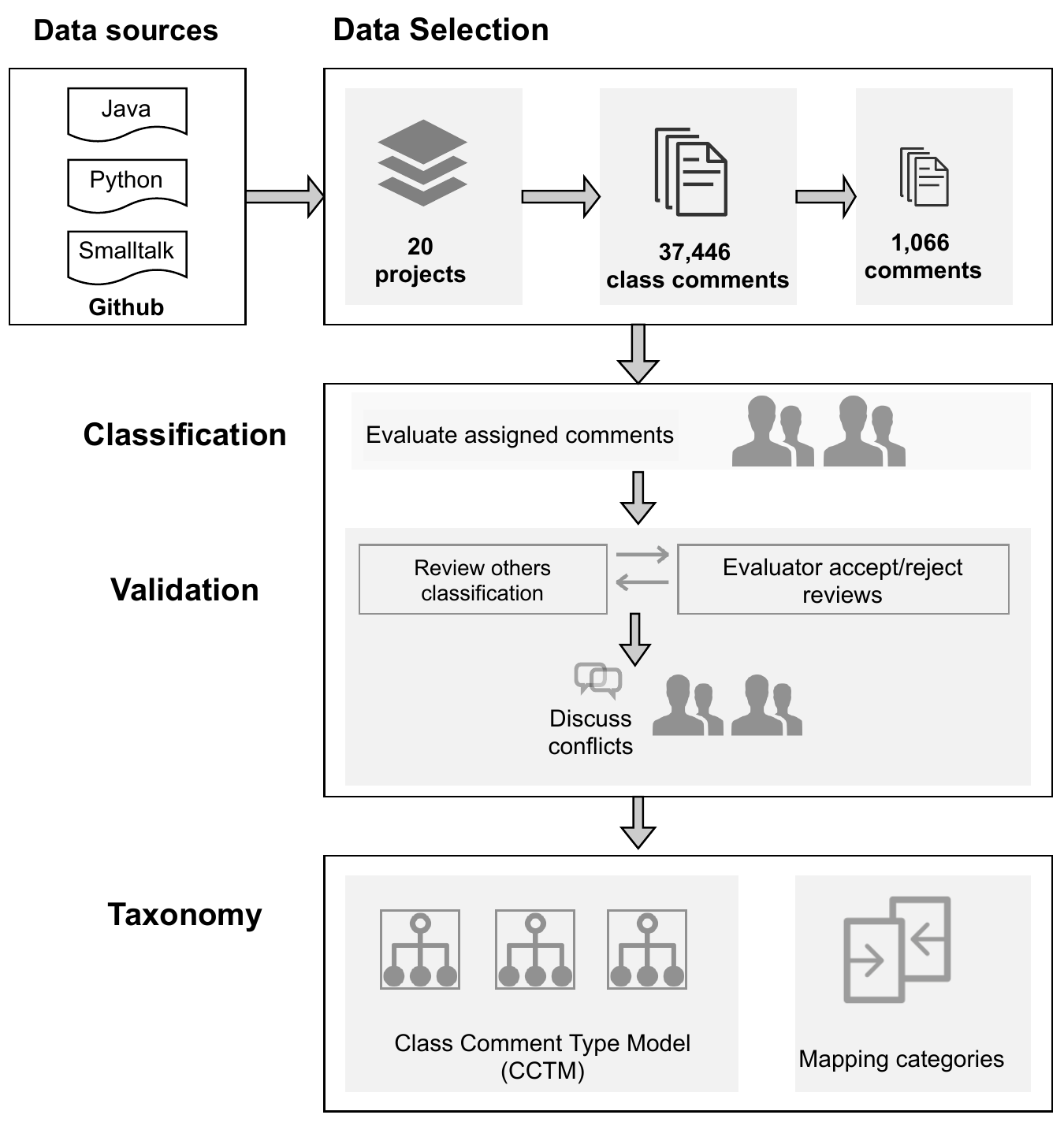}
    \caption{Taxonomy Study (RQ$_1$)}
    \figlabel{RQ1pipeline}
        \vspace{-4mm}
\end{figure}

\figref{RQ1pipeline} depicts the research approach followed to answer RQ$_1$. 
The outcome of this research, as explained later, consists of a mapping taxonomy and a comprehensive taxonomy of class comment types, called \emph{CCTM} (\emph{\textbf{C}lass \textbf{C}omment \textbf{T}ype \textbf{M}odel}), mined from the actual commenting practices of developers (see \figref{RQ1pipeline}). 

\textbf{Mapping categories:}
Before preparing the \emph{CCTM}, we analyzed various earlier comment taxonomies, such as the
code comment taxonomy for Java~\cite{Pasc17a} and Python~\cite{Zhan18a}, and the class comment taxonomy for Smalltalk~\cite{Rani21b}, to analyze their reported categories.
We mapped the categories from each existing taxonomy, since they were formulated using different approaches and based on different comment scope and categories.
For instance, the Python code comment taxonomy by Zhang~\cite{Zhan18a} is inspired by the Java code comment taxonomy~\cite{Pasc17a}, whereas the Smalltalk class comment taxonomy is formulated using an open-card sorting technique~\cite{Rani21b}.
Given the importance of mapping categories from heterogeneous environments~\cite{Choi06a}, we established semantic interoperability~\cite{Euze01a} of the categories from each taxonomy in this step.
One evaluator mapped the categories from Smalltalk to Java and Python categories.
Two other evaluators validated the mapping by reviewing each mapping and proposing the changes.
The original evaluator accepted or rejected the changes.
All the disagreement cases were reviewed by the fourth evaluator and discussed among all to reach the consensus.

The categories that did not map to other taxonomy, we added them as new categories in that taxonomy. For example, the \emph{Precondition} category from Smalltalk did not map to Java and Python, and thus we added it as a new category in Java and Python. Thus, we proposed the \emph{CCTM} taxonomy highlighting the existing and new categories for class comments in the next step \emph{Preparing CCTM}.

\textbf{Preparing CCTM:} In this step, we analyzed various ecosystems from each language.
We built our dataset\footnote{\fontsize{6}{6}\selectfont \repFolder{Dataset}} by mining data from 20 GitHub  projects (see \secref{data-collection} for further details).
In the \emph{Data Selection} step, we extracted the classes and their comments, collecting a total of 37\,446 class comments.
We selected a statistically significant sample set for each language summing 1\,066 total class comments. We then qualitatively classified the selected class comments (see the following \emph{Classification} step) and validated them (see following \emph{Validation} step) by reviewing and refining the categorization. 

\textbf{Classification:} Four evaluators (two Ph.D.\ candidates, and two faculty members, all authors of this paper) each having at least four years of programming experience, participated in the study.
We partitioned Java, Python, and Smalltalk comments equally among all evaluators based on the distribution of the language's dataset to ensure the inclusion of comments from all projects and diversified lengths.
Each evaluator classified the assigned class comments according to the CCTM taxonomy of Java, Python and Smalltalk~\cite{Pasc17a,Zhan18a,Rani21b}.
\begin{figure}[th]
    \centering
    \includegraphics[width=\linewidth,height=0.65\linewidth]{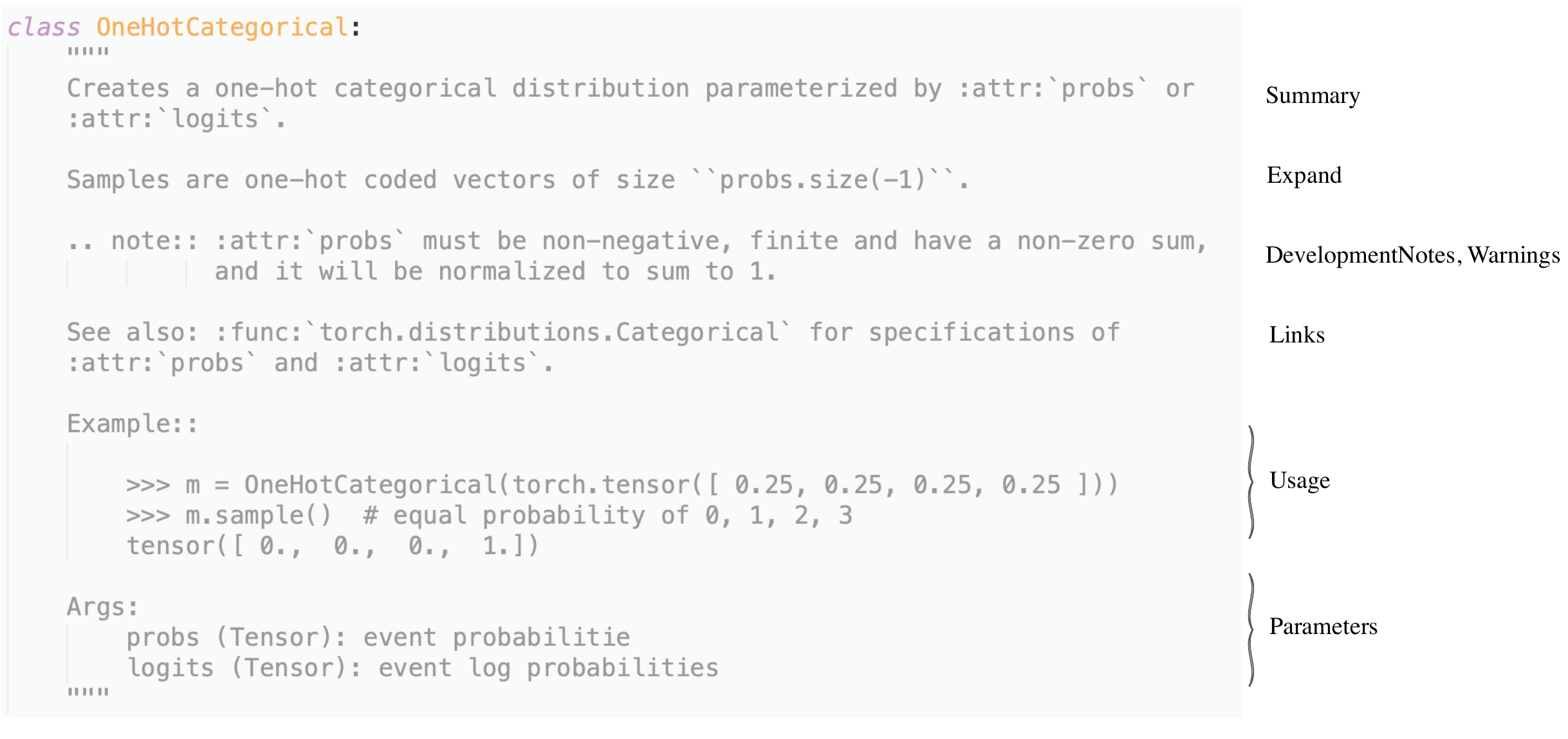}
    \caption{Class comment of Python (\figref{python-class-comment}) classified in various categories}
    \figlabel{python-class-comment-classified}
    \vspace{-1mm}
\end{figure}

For example, the Python class comment in \figref{python-class-comment} is classified in the categories such as \emph{Summary}, \emph{Warnings}, and \emph{Parameters} \etc, as shown in \figref{python-class-comment-classified}.
\textbf{Validation:}
After completing their individual evaluations, the evaluators continued with the validation step.
Thus, the evaluators adopted a three-step method to validate the correctness of the performed class comments classification.
In the first iteration, called \emph{``Review others' classification''} in \figref{RQ1pipeline}, every evaluator was tasked to review 50\%  of the comments (randomly assigned and) classified by other evaluators.
This step allowed us to confirm that each evaluator's classification is checked by at least one of the other evaluators.
In reviewing the classifications, the reviewers indicated their judgement by labeling each comment with \emph{agree} or \emph{disagree} labels.

In the second iteration (called \emph{``Evaluator accept or reject reviews''} in \figref{RQ1pipeline}), the original evaluator examined the disagreements and the proposed changes.
They indicated their opinion for the changes by accepting the change or rejecting it, stating the reason.
In case the reviewer's changes were accepted, the classification was directly fixed, otherwise the disagreements were carried to the next iteration.
The third iteration assigned all identified disagreements for review to a new evaluator, who had not yet looked at the classification.
Based on a majority voting mechanism, a decision was made and the classification was fixed according to agreed changes.
The levels of \emph{agreement and disagreement} among the evaluators for each project and language are available in the replication package.\footnote{\repFolder{Result/RQ1/Manually-classified-comments}}

After arriving at a decision on all comment classifications, we merged overlapping categories or renamed the classes by  applying the majority voting mechanism, thus converging on a final version of the taxonomy, \ie CCTM.
This way, all the categories were discussed by all the evaluators to select the best naming convention, and whenever required, unnecessary categories were removed and duplicates were merged.

\emph{RQ$_2$: Automated Classification of Class Comment Types in Different Programming Languages}\\

\textbf{Motivation.} Previous work has focused on identifying information types from code comments scattered throughout the source code, from high-level class overviews to low-level method details~\cite{Stei13b,Pasc17a,Zhan18a,Geis20a}.
These works have focused individually on code comments in Java, Python, C++, and COBOL.
Differently from our study, none of these works attempted to identify information types in class comments automatically across multiple languages. 
In our work, we are interested in exploring strategies that are able to achieve this goal in multiple languages such as Python, Java and Smalltalk. 
Hence, for this research question, we explore to what extent term-based strategies and techniques based on NLP patterns \cite{Sorb19a,Sorb16a,Pani15a} help to automatically identify the different information types composing CCTM.

\textbf{Automated classification of class comment types}. After the definition of CCTM, we propose an approach, called \texttt{TESSERACT} (au\textbf{T}omated multi-language\textbf{E} cla\textbf{SS}ifi\textbf{ER} of cl\textbf{A}ss \textbf{C}ommen\textbf{T}s), which automatically classifies  class comment according to CCTM. 
To achieve this goal, we considered all the comments manually validated in answering RQ$_1$.
Specifically, our approach leverages machine learning (ML) techniques and consists of four main steps:

\begin{enumerate}

\item \textbf{\emph{Preprocessing}}: 
All the manually-labeled class comments from RQ$_1$ in our dataset are used as ground truth to classify the unseen class comments.\footnote{\repFile{Dataset/RQ2/Java/raw.csv}}
It is important to mention that, since the classification is sentence-based, we split the comments into sentences.
As a common preprocessing step, we change the sentences to lower case and remove all special characters.\footnote{\repFile{Results/RQ2/All-steps-result.sqlite}}
We apply typical preprocessing steps on sentences \cite{Baez99b} (\eg stop-word removal) for TEXT features but not for NLP features to preserve the word order to capture the important n-gram patterns observed in the class comments~\cite{Rani21b}. 
\begin{figure}[th]
    \centering
    \includegraphics[width=\linewidth]{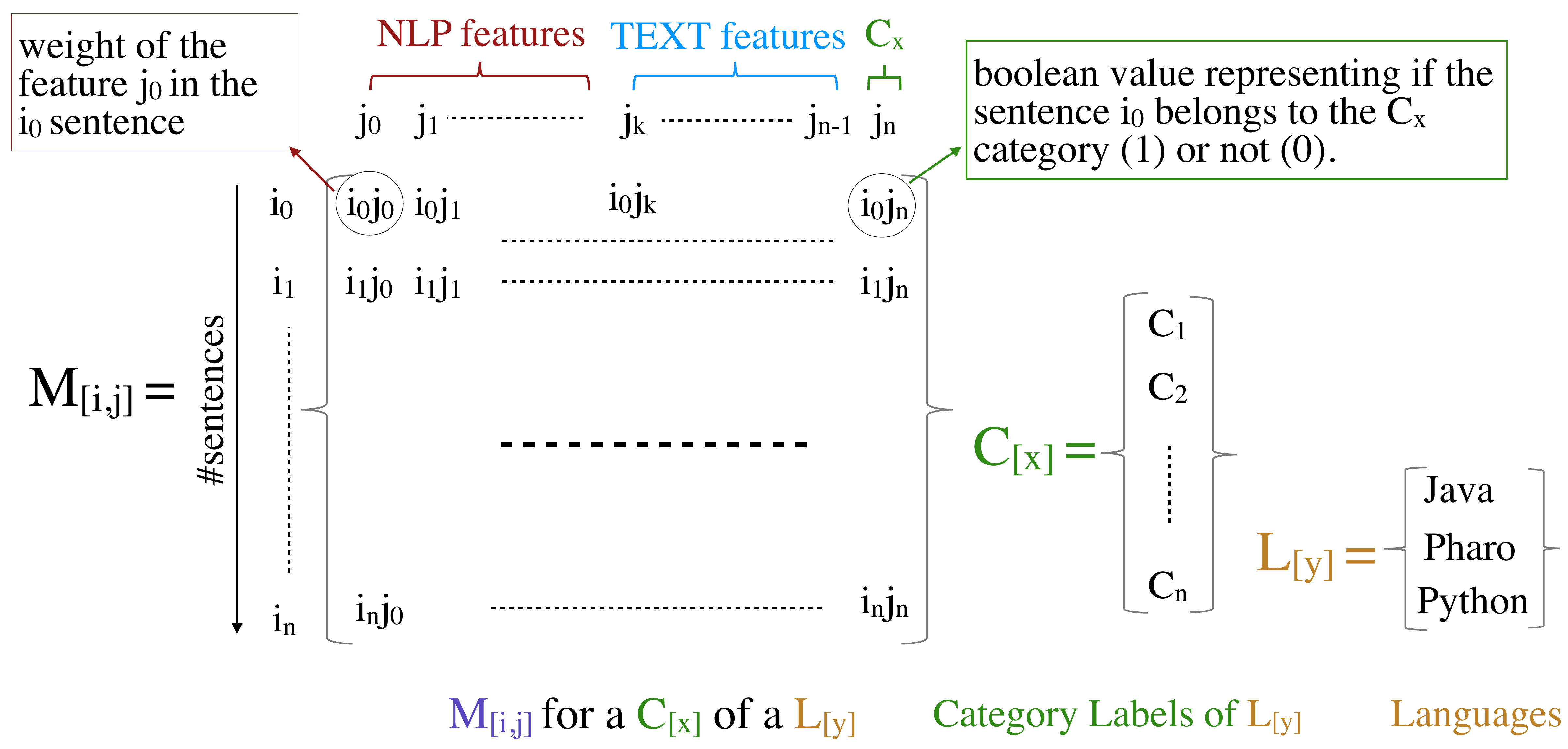}
    \caption{Matrix representation for a classifier}
    \figlabel{classifier-matrix}
        \vspace{-4mm}
\end{figure}

\item \textbf{\emph{NLP Feature Extraction}}:
In this phase we focus on extracting NLP features to add to 
an initial term-by-document matrix \emph{M} shown in \figref{classifier-matrix} where each row represents a \textit{comment sentence} (\ie a sentence belongs to our language dataset composing CCTM) and each column represents the extracted feature. 
To extract the NLP features, we use NEON, a tool proposed in previous work \cite{Sorb19a}, which is able to automatically detect NLP patterns (\ie predicate-argument structures recurrently used for specific intents~\cite{Sorb15a}) present in natural language descriptions composing various types of software artifacts (\eg mobile user reviews, emails \etc)~\cite{Sorb19a}. 
In our study, we use NEON to infer all NLP patterns characterizing the comment sentences modeled in the matrix \emph{M}.
Then, we add the identified NLP patterns as feature columns in \emph{M}, where each of them models the presence or absence (using binomial features) of an NLP pattern in the comment sentences.
More formally, the boolean presence or absence of a j$-th$ NLP pattern (or feature) in a generic i$-th$ sentence in \emph{M} is modeled by 0 (absence) and 1 (presence) values, respectively.
The output of this phase consists of the matrix \emph{M} where each i$-th$ row represents a comment sentence and j$-th$ represents an NLP feature.

\item  \textbf{\emph{TEXT Features}}:
In this phase, we add additional features (TEXT features) to the matrix \emph{M}.
To get the TEXT features, we preprocess the comment sentences by applying stop-word removal\footnote{\url{http://www.cs.cmu.edu/~mccallum/bow/rainbow/}} and stemming~\cite{Lovi68a}.\footnote{\url{https://weka.sourceforge.io/doc.stable/weka/core/stemmers/IteratedLovinsStemmer.html}}
The output of this phase corresponds to the matrix \emph{M} where each row represents a \textit{comment sentence} (\ie a sentence belonging to our language dataset composing CCTM) and each column represents a term contained in it. 
More formally, each entry \textbf{M$_{[i,j]}$} of the matrix represents the weight (or importance) of the j$-th$ term contained in the i$-th$ \textit{comment sentence}. 

For the TEXT features, terms in \emph{M} are weighted using the \emph{tf-idf} score \cite{Baez99b}, which identifies the most important terms in the sentences in matrix \emph{M}. 
In particular, we use \emph{tf-idf} as it downscales the weights of frequent terms that appear in many sentences. 
Such a weighting scheme has been successfully adopted in recent work~\cite{Misr20a} for performing code comment classification.
The output of this phase consists of the weighted matrix \emph{M} where each row represents a comment sentence and a column represents the weighted term contained in it. 

It is important to note that a generic i$-th$ comment sentence can be classified into multiple categories according to CCTM.
To model this, we prepare the matrix \emph{M} for each category of each language. The generic (last) column $\textbf{M}[j_n]$ of the matrix \textbf{M}  (where $n-1$ is the total number of features extracted from all sentences) represents the category $\textbf{C}_{[x]}$ of a language $\textbf{L}_{[y]}$ as shown in \figref{classifier-matrix}.
More formally, each entry \textbf{M$_{[i,j_n]}$} of the matrix represents the boolean value if the i$-th$ sentence belongs to the $\emph{C}_{x}$ (1) or not(0).

\item \textbf{\emph{Classification}}:
We automatically classify class comments by adopting various machine learning models and a 10-fold cross-validation strategy. 
These machine learning models are fed with the aforementioned matrix \emph{M}.
Specifically, to increase the generalizability of our findings, we experiment (relying on the Weka tool\footnote{\url{http://waikato.github.io/weka/}}) with several machine learning techniques, namely, the standard probabilistic Naive Bayes classifier, the J48 tree model, and the Random Forest model.
It is important to note that the choice of these techniques is not random but based on their successful usage in recent work on code comment analysis~\cite{Stei13b,Pasc17a,Zhan18a,Shin18a} and classification of unstructured texts for software maintenance purposes \cite{Pani15a,Sorb16a}.
\end{enumerate}

\textbf{Evaluation Metrics \& Statistical Tests}. To evaluate the performance of the tested ML techniques, we adopt well-known information retrieval metrics, namely  precision, recall, and F-measure~\cite{Baez99b}. 
During our empirical investigation, we focus on investigating the best configuration of features and machine learning models as well as alleviating concerns related to overfitting and selection bias. 
Specifically,
(i) we investigate the classification results of the aforementioned machine learning models with different combinations of features (NLP, TEXT, and TEXT+NLP features), by adopting a 10-fold validation strategy on the term-by-document matrix \emph{M}; 
(ii) to avoid potential bias or overfitting problems, we train the model for the categories having at least 40 manually validated instances in our dataset.
The average number of comments belonging to a category varies from 43 comments to 46 comments across all languages, therefore we selected the categories with a minimum of 40 comment instances.
The top categories selected from each language with the number of comments are shown in \tabref{all_projects_top_categories}.
In order to determine whether the differences between the different input features and classifiers were statistically significant or not we performed a Friedman test, followed by a post-hoc Nemenyi test, as recommended by Dem{\v s}ar~\cite{Dems06a}.
Results concerning RQ$_2$ are reported in \secref{results}.

\begin{table}[ht]
    \small
    \caption{Top frequent categories with at least 50 comments}
    \tablabel{all_projects_top_categories}
    \begin{tabular}{lll}
    \noalign{\smallskip}\hline
    \textbf{Language} &  \textbf{Categories} & \textbf{\#Comments} \\
    \noalign{\smallskip}\hline
        Java    & Summary   & 336 \\
                & Expand    & 108 \\
                & Ownership & 97 \\
                & Pointer   & 88 \\
                & Usage     & 87 \\
                & Deprecation & 84 \\
                & Rationale  & 50 \\
        \hline
        Python  & Summary   & 318 \\
                & Usage     & 92 \\
                & Expand    & 87 \\
                & Development Notes & 67 \\
                & Parameters    & 57 \\  
        \hline
        Smalltalk  & Responsibility	& 240 \\
                & Intent	& 190 \\
                & Collaborator	& 91 \\
                & Examples	& 81 \\
                & Class Reference 	& 57 \\ 
                & Key Message & 48 \\
                & Key Implementation Point & 46 \\
        \noalign{\smallskip}\hline
    \end{tabular}
\end{table}

\section{Results}
\seclabel{results}

This section discusses the results of our empirical study.

\subsection{\texorpdfstring{RQ$_1$}: Class Comment Categories in Different Programming Languages}
\seclabel{resultsRq1}

\begin{figure*}[ht]
    \centering
    \includegraphics[width=\linewidth]{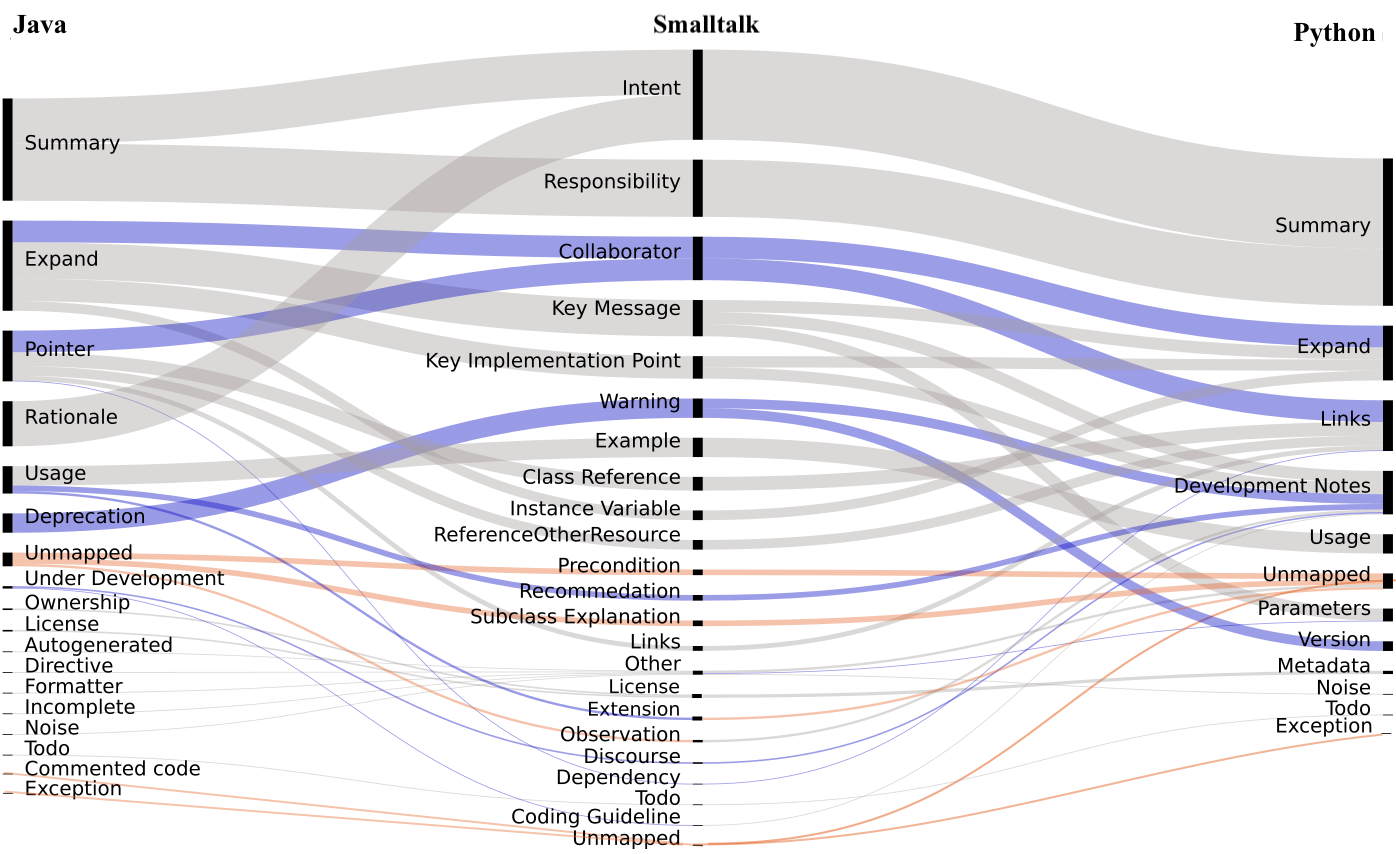}
    \caption{Mapping of Smalltalk categories to Java and Python}
    \figlabel{mapping-pharo-java-python}
        \vspace{-4mm}
\end{figure*}
As a first step to formulating the CCTM taxonomy we systematically map the available taxonomies from previous works and identify the unmapped categories as shown in \figref{mapping-pharo-java-python}.
The mapping taxonomy shows a number of cases in which Java and Python taxonomies do not entirely fit the Smalltalk taxonomy, as shown by pink and violet edges in \figref{mapping-pharo-java-python}.
The figure shows the information types particular to a language (highlighted with red edges and unmapped nodes) such as \emph{Subclass Explanations}, \emph{Observation}, \emph{Precondition}, and \emph{Extension} are found in the Smalltalk taxonomy but not in Python and Java.
However, our analysis shows that these information types are present in the class comments of Java and Python projects. 
We introduce such categories to the existing taxonomies of Java and Python, and highlight them in green in \figref{info-types-all-projects}. 
On the other hand, the categories such as \emph{Commented code}, \emph{Exception}, and \emph{Version} are found in Java and Python class comments but not in Smalltalk.
One reason can be that the commented code is generally found in inline comments instead of documentation comments. 
However, information about \emph{Exception} and \emph{Version} is found in class comments of Java and Python but not in Smalltalk.

The mapping taxonomy also highlights the cases where categories from different taxonomies match partially.
We define such categories as subset categories and highlight them with violet edges.
For example, the \emph{Deprecation} category in Java and the \emph{Version} category in Python are found under the \emph{Warning} category in Smalltalk but their description given in the respective earlier work covers only a subset of that information type according to our investigation.
Pascarella \etal define the \emph{Deprecation} category as ``\emph{it contains explicit warnings used to inform the users about deprecated interface artifacts. The tag comment such as @version, @deprecated, or @since is used}'' whereas Zhang \etal define the category \emph{Version} as ``\emph{identifies the applicable version of some libraries}'' but do not mention the deprecation information in this category or any other category in their taxonomy~\cite{Pasc17a,Zhan18a}.
Thus, we define the \emph{Version} category as a subset or a partial match of the \emph{Deprecation} category.
On the other hand, the \emph{Warning} category in Smalltalk (``\emph{Warn readers about various implementation details of the class}'') covers a broader aspect of warnings than the \emph{Deprecation} category but does not mention the \emph{Version} information type~\cite{Rani21b}.
We mark these categories as subset categories, and highlight them with violet edges.
Similarly, the \emph{Collaborator} category in Smalltalk matches partially \emph{Expand} and \emph{Links} in Python.
The categories such as \emph{Expand}, \emph{Links}, and \emph{Development Notes} in Python combine several types of information under them compared to Smalltalk.
For example, \emph{Expand} includes collaborators of a class, key methods, instance variables, and implementation-specific details.  
Such categories formulate challenges in identifying a particular type of information from the comment.

\finding{1}{The Python taxonomy focuses more on high-level categories, combining various types of information into each category whereas the Smalltalk taxonomy is more specific to the information types.}


Using the categories from the mapping taxonomy, we analyzed class comments of various languages and formulated the taxonomy for each language to answer the RQ$_1$.
\begin{figure}[t]
    \centering
    \includegraphics[width=\linewidth]{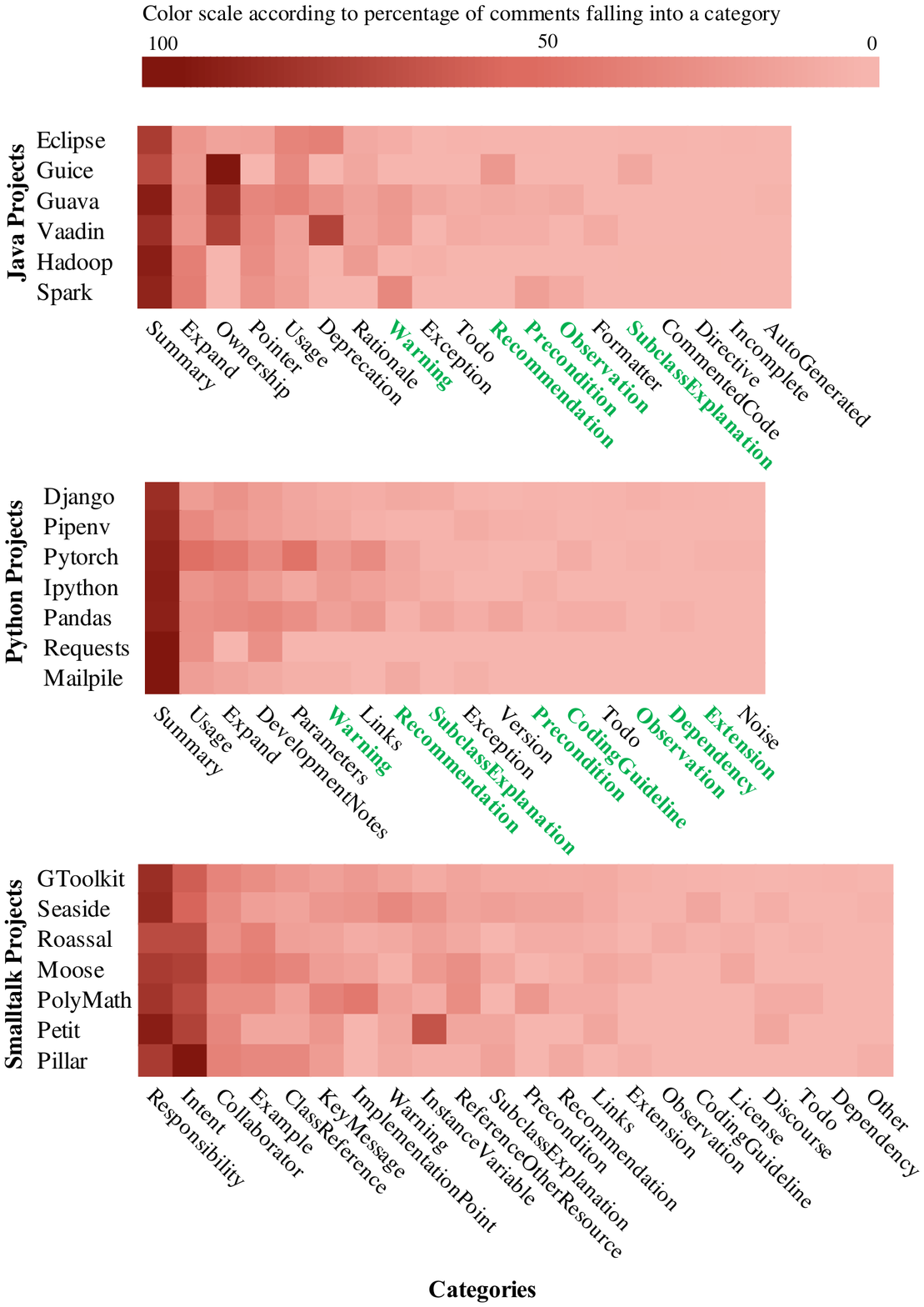}
    \caption{Categories found in class comments (\emph{CCTM}) of various projects (shown on the y-axis) of each programming language. The x-axis shows the categories inspired from existing work (highlighted in black) and the new categories (highlighted in green).)}
    \figlabel{info-types-all-projects}
        \vspace{-4mm}
\end{figure}
\figref{info-types-all-projects} shows the frequency of information types per language per project.
The categories shown in green are the newly-added categories in each taxonomy.
The categories in each heatmap are sorted according to the frequency of their occurrence in total.
For example, in Java, \emph{Summary} appeared in a total of 336 comments (88\%) of six Java projects out of 378 comments.
Pascarella~\cite{Pasc17a} proposed a hierarchical taxonomy, grouping the lower-level categories within the higher-level categories such as grouping the categories \emph{Summary}, \emph{Rationale}, and \emph{Expand} within \emph{Purpose}.
We show only lower-level categories that correspond with identified information types from other languages.

\figref{info-types-all-projects} shows that a few types of information, such as the summary of the class (\emph{Summary}, \emph{Intent}), the responsibility of the class (\emph{Responsibility}, \emph{Summary}), links to other classes or sources (\emph{Pointer}, \emph{Collaborator}, \emph{Links}), developer notes (\emph{Todo}, \emph{Development Notes}), and warnings about the class (\emph{Warning}) are found in class comments of all languages.
Summary being the most prevalent category in all languages affirms the value of investing effort in summarizing the classes automatically~\cite{Haid10a,Naza16a,Bink13a,More13c,Drag10a}. 
As a majority of summarization techniques focus on generating the intent and responsibilities of the class for program comprehension tasks~\cite{More13c}, other information types such as \emph{Warning}, \emph{Recommendation}, and \emph{Usage} are generally ignored even though coding style guidelines suggest to write them in documentation comments.
Our results indicate that developers mention them frequently, but whether they find these information types important to support specific development tasks, or they write just to adhere to the coding guidelines,  requires more analysis. Such information types present an interesting aspect to investigate in future work.
For example, usage of a class (\emph{Usage}), its key responsibilities (\emph{Responsibility}), warnings about it (\emph{Warning}), and its collaborators (\emph{Collaborator}) are found in significant numbers of comments in all languages. These information types are often suggested by the coding guidelines as they can support developers in various development and maintenance tasks.
These information types can be included in the customized code summaries based on the development tasks a developer is doing.
For example, a developer seeking dependant classes can quickly find such classes from the class comment without reading the whole comment. 
Similarly, a developer expected to refactor a legacy class can quickly go through the warnings, if present, to understand the specific conditions better and thus can save time. 
We plan to investigate such information types with developers. Specifically, we are interested in exploring how various categories are useful to developers, and for which kinds of tasks \eg program comprehension tasks or maintenance tasks.

\finding{2}{Developers embed various types of information in class comments, varying from the high-level overview of the class to the low-level implementation details of the class across the investigated languages.}

According to Nurvitadhi \etal~\cite{Nurv03a}, a class comment in Java should describe the purpose of the class, its responsibilities, and {its} interactions with other classes.
In our study, we observe that class comments in Java often contain the purpose and responsibilities of the class (\emph{Summary} and \emph{Expand}), but its interactions with other classes (\emph{Pointer}) less often. 
On the other hand in Smalltalk, the information about interactions with other classes, \ie \emph{Collaborator}, is the third most frequent information type after \emph{Intent} and \emph{Responsibility} compared to Java and Python.
One of the reasons can be that Smalltalk class comments are guided by a CRC (Class, Responsibility, Collaborator) design template and developers follow the template in writing {these information types}~\cite {Rani21b}.
Class comments in Java also contain many other types of information.
The most frequent type of information present in class comments is \emph{Summary}, which shows that developers summarize the classes in the majority of the cases. 
Pascarella \etal\ found \emph{Usage} to be the second most prevalent category in code comments, while we find overall \emph{Expand} to be the second most prevalent category in class comments, and \emph{Usage} to be the fifth most prevalent type of information~\cite{Pasc17a}.
However, the {most} prevalent categories vary across projects of a language, and also across programming languages.
For example, \emph{Usage} is mentioned more often than \emph{Expand} in Google projects (Guice and Guava) whereas in Apache projects (Spark, Hadoop) it is not.
In contrast to Java, Python class comments contain
\emph{Expand} and \emph{Usage} equally frequently thus showing that Python targets both end-user developers and internal developers.
We notice that Python and Smalltalk class comments contain more low-level implementation details about the class compared to Java.
For example, Python class comments contain the details about the class attributes and the instance variables of a class with a header ``\emph{Attributes}'' or ``\emph{Parameters}'', its public methods with a header ``\emph{Methods}'', and its constructor arguments in the \emph{Parameters} and \emph{Expand} categories.
Additionally, Python class comments often contain explicit warnings about the class (with a header ``\emph{warning:}'' or ``\emph{note:}'' in the new line), making the information easily noticeable whereas such behavior is rarely observed in Java.
Whether such variations in the categories across projects and languages are due to different project comment style guidelines or due to developer personal preferences is not known yet.
We observe that developers use common natural language patterns to write similar types of information.
For example, a Smalltalk developer described the collaborator class of the ``PMBernoulliGeneratorTest'' class in \autoref{lst:Collaborator-PMBernoulliGeneratorTest-class} and a Java developer as shown in \autoref{lst:Collaborator-SequenceFileRecordReader-class} thus showing a pattern \emph{``[This class] for [other class]''} to describe the collaborating classes.
This information type is captured in the categories \emph{Collaborator} in Smalltalk and \emph{Pointer} in Java.
\begin{minipage}{\linewidth}
    \vspace{2mm}
    \begin{lstlisting}[caption={Collaborator mentioned in the PMBernoulliGeneratorTest class in Smalltalk}, label={lst:Collaborator-PMBernoulliGeneratorTest-class}]
A BernoulliGeneratorTest is a test class for testing the behavior of BernoulliGenerator
\end{lstlisting}
    \end{minipage}
    \vspace{1mm}
\begin{minipage}{\linewidth}
    \begin{lstlisting}[caption={Collaborator mentioned in the SequenceFileRecordReader class in Java}, label={lst:Collaborator-SequenceFileRecordReader-class}]
An {@link RecordReader} for {@link SequenceFile}s.
\end{lstlisting}
    \end{minipage}

Identifying such patterns can help in extracting the type of information from the comment easily, and can support the developer by highlighting the required information necessary for a particular task, \eg to modify the dependent classes in a maintenance task.

In contrast to earlier studies, we observe that developers mention details of their subclasses in a parent class comment in all languages.
We group this information under the \emph{Subclass Explanation} category.
In the Javadoc guidelines, this information is generally indicated by a special \code{@inherit} tag in the method comments, but we did not find such a guideline for Java class comments.
Similarly we found no such guideline to describe subclasses for Smalltalk class comments or method comments.
In contrast, the standard Python style guideline~\cite{Pyth20a} suggests adding this information in the class comment but other Python style guidelines such as those from Google\footnote{\url{https://sphinxcontrib-napoleon.readthedocs.io/en/latest/example_google.html}} and Numpy\footnote{\url{https://numpydoc.readthedocs.io/en/latest/format.html}} do not mention this information type.
However, we find instances of class comments containing subclass information in  the \emph{IPython} and \emph{Pytorch} projects that follow the Numpy and Google style guidelines respectively.
Investigating which information types each project style guidelines suggest for the comments, and to what extent developers follow these style guidelines in writing class comments is future work.

\finding{3}{Not all information types found in class comments are suggested by corresponding project style guidelines.}


\textbf{\emph{Discussion:}}
In the process of validating the taxonomies, reviewers (evaluator reviewing the classification) marked their disagreement for the classification, stating their reason, and proposing the changes.
A majority of disagreements in the first Smalltalk iteration were due to the long sentences containing different types of information (\emph{Intent} and \emph{Responsibility} information types were commonly interweaved), and assignment of information to categories \emph{Key Implementation Point} and \emph{Collaborator}.
In Java and Python, we observed that disagreements were due to the broad and loosely defined categories such as \emph{Expand} in Java and \emph{Development Notes} in Python.
Several information types such as \emph{Warning}, \emph{Recommendation}, \emph{Observation}, and \emph{Development Notes} are not structured by special tags and thus pose a challenge for automatic identification and extraction.
On the other hand, a few categories such as \emph{Example} and \emph{Instance Variable} in Smalltalk, and \emph{Deprecation}, \emph{Links}, and \emph{Parameters} in Java and Python are explicitly marked by the headers or the tags such as \emph{Usage}, \emph{Instance Variable}, \emph{@since}, \emph{@see}, \emph{@params} respectively.
We observed that developers use common keywords across languages to indicate a particular information type.
For example, notes are mentioned in the comment with a keyword ``note'' as a header as shown in \lstref{Explicit-note-Resources-class}, \lstref{Explicit-note-Conv3d-class}, and \lstref{Explicit-note-BlElementPositionChangedEvent-class}.
\begin{minipage}{\linewidth}
    \vspace{2mm}
\begin{lstlisting}[caption= {Explicit note mentioned in the Resources class in Java}, label={lst:Explicit-note-Resources-class}]
Note that even though these 
* methods use {@link URL} parameters, they are usually not appropriate for HTTP or other
* non-classpath resources.
\end{lstlisting}
\end{minipage}
\begin{minipage}{\linewidth}
    \begin{lstlisting}[caption= {Explicit note mentioned in the Conv3d class in Python}, label={lst:Explicit-note-Conv3d-class}]
.. note::
    
Depending of the size of your kernel, several (of the last)
columns of the input might be lost, because it is a valid `cross-correlation`_,
and not a full `cross-correlation`_.
It is up to the user to add proper padding.
\end{lstlisting}
\end{minipage}
\begin{minipage}{\linewidth}
    \begin{lstlisting}[caption= {Explicit note mentioned in the BlElementPositionChangedEvent class in Smalltalk}, label={lst:Explicit-note-BlElementPositionChangedEvent-class}]
Note: position may change even if an element has no parent\end{lstlisting}
\end{minipage}

Several information types are suggested by the project-specific style guidelines but not the exact syntax whereas a large number of information types are not mentioned by them.
Due to the lack of conventions for these information types, developers use their own conventions to write them in the comments. 

Maalej \etal~\cite{Maal14a} demonstrates that developers consult comments in order to answer their questions regarding program comprehension. 
However, different types of information are interweaved in class comments and not all developers need to know all types of information.
Cioch \etal presented the documentation information needs of developers depending on the stages of expertise\cite{Cioc96a}.
They showed that experts need design details and low-level details whereas novice developers require a high-level overview with examples.
Therefore, identifying and separating these information types is essential to address the documentation needs.
Rajlich presented a tool that gathers important information such as a class's responsibilities, its dependencies, member functions, and authors' comments to facilitate the developer's need to access the particular information types~\cite{Rajl00b}.
We advance the work by identifying and extracting several other frequent information types from the class comments automatically.


\subsection{RQ$_2$: Automated Classification of Class Comment Categories in Different Programming Languages}
\seclabel{resultsRq2}
Haiduc \etal~\cite{Haid10a} performed a study on automatically generating summaries for classes and methods and found that the experimented summarization techniques work better on methods than classes. 
More specifically, they discovered that while developers generally agree on the important attributes that should be considered in the method summaries, there were conflicts concerning the types of information (\ie class attributes and method names) that should appear in the class summaries.
In our study, we found that while Smalltalk and Python developers frequently embed class attributes or method names in class comments, it rarely happens in Java. 
Automatically identifying various kinds of information from comments can enable the generation of customized summaries based on what information individual developers consider relevant for the task at hand (\eg maintenance task). 
To this aim, as described in \secref{analysis-method}, we empirically experiment with a machine learning-based multi-language approach to automatically recognize the types of information available in class comments. 

\begin{table*}[ht]
    \centering
    \small
    \captionsetup{font=small}
\caption{Results for Java, Phyton, and Smalltalk obtained through different machine learning models and features}
\tablabel{combination_results}
\begin{tabular}{|c|l|c|c|c|c|c|c|c|c|c|}
\hline
\multirow{2}{*}{\textbf{Language}} & \multicolumn{1}{c|}{\multirow{2}{*}{\textbf{ML Models}}} & \multicolumn{3}{c|}{\textbf{TEXT}}                        & \multicolumn{3}{c|}{\textbf{NLP}}                         & \multicolumn{3}{c|}{\textbf{NLP + TEXT}}                  \\ \cline{3-11} 
                                   & \multicolumn{1}{c|}{}                                    & \emph{Precision} & \emph{Recall} & \emph{F-Measure} & \emph{Precision} & \emph{Recall} & \emph{F-Measure} & \emph{Precision} & \emph{Recall} & \emph{F-Measure} \\ \hline
\multirow{3}{*}{Java}              & J48  
                                        & \textbf{0.91}      & \textbf{0.92}   & \textbf{0.91}      
                                        & \textbf{0.84}      & 0.81            & 0.81               
                                        & 0.89               & 0.90            & 0.88               \\ \cline{2-11} 
                                   & Naive Bayes   
                                        & 0.86               & 0.83            & 0.83               
                                        & \textbf{0.84}      & 0.81            & 0.81               
                                        & 0.86               & 0.84            & 0.84               \\ \cline{2-11} 
                                    & Random Forest                                            
                                        & \textbf{0.91}      & \textbf{0.92}   & \textbf{0.91}      
                                        & \textbf{0.84}      & \textbf{0.87}   & \textbf{0.82}      
                                        & \textbf{0.92}      & \textbf{0.92}   & \textbf{0.92}      \\ \hline
\multirow{3}{*}{Python}            & J48  
                                        & 0.73               & 0.83            & 0.73               
                                        & 0.68               & 0.80            & 0.66               
                                        & 0.81               & 0.83            & 0.79               \\ \cline{2-11} 
                                   & Naive Bayes  
                                        & 0.78               & 0.69            & 0.72               
                                        & 0.75               & 0.77            & 0.75               
                                        & 0.79               & 0.72            & 0.74               \\ \cline{2-11} 
                                   & Random Forest   
                                        & \textbf{0.84}      & \textbf{0.85}   & \textbf{0.83} 
                                        & \textbf{0.78}      & \textbf{0.81}   & \textbf{0.78}     
                                        & \textbf{0.85}      & \textbf{0.86}   & \textbf{0.84}      \\ \hline
\multirow{3}{*}{Smalltalk}         & J48                                                      
                                        & 0.60               & 0.87            & 0.58               
                                        & 0.61               & \textbf{0.88}   & 0.59               
                                        & 0.72               & 0.88            & 0.70               \\ \cline{2-11} 
                                   & Naive Bayes 
                                        & \textbf{0.84}      & 0.80            & \textbf{0.82}      
                                        & \textbf{0.85}      & 0.83            & \textbf{0.83}      
                                        & \textbf{0.86}      & 0.82            & \textbf{0.84}      \\ \cline{2-11} 
                                   & Random Forest 
                                        & 0.75               & \textbf{0.90}   & 0.73               
                                        & 0.82               & \textbf{0.88}   & 0.82               
                                        & 0.78               & \textbf{0.90}   & 0.77               \\ \hline
\end{tabular}
\end{table*}

\begin{figure*}[ht]
    \centering
    \includegraphics[width=\linewidth]{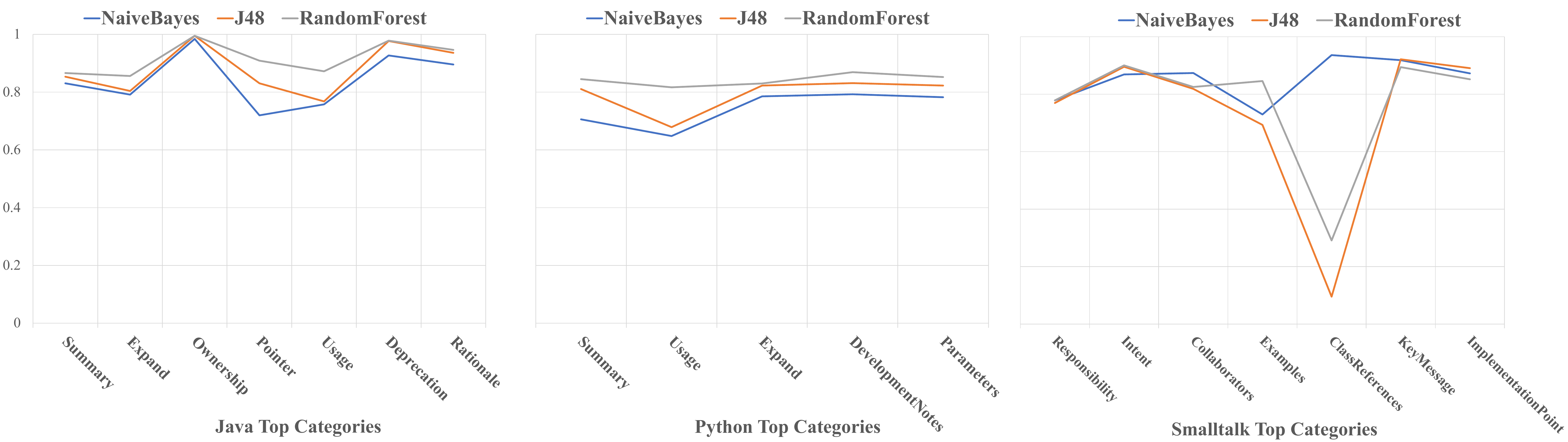}
    \caption{Performance of the different classifiers based on F-measure for the TEXT + NLP feature set}
    \figlabel{classifiers-comparison-text-nlp}
        \vspace{-4mm}
\end{figure*}

\tabref{combination_results} provides an overview of the average precision, recall, and F-Measure results considering the (top frequent) categories for all languages shown in \tabref{all_projects_top_categories}.
The results are obtained using multiple machine learning models and various combination of features\footnote{\repFile{Result/RQ2/CV-10-results.xlsx}}: (i) TEXT features only, (ii) NLP features only, (iii) both NLP and TEXT features.\footnote{\repFile{Result/RQ2/All-steps-results.sqlite}}
The results in \tabref{combination_results} show that the NLP+TEXT configuration achieves the best results with the Random Forest algorithm with relatively high precision (ranging from 78\% to 92\% for the selected languages), recall (ranging from 86\% to 92\%), and F-Measure (ranging from 77\% to 92\%). 
\figref{classifiers-comparison-text-nlp} shows the performance of the different algorithms with NLP+TEXT features for the most frequent categories of each language.

\finding{4}{Our results suggest that the Random Forest algorithm fed by the combination of NLP+TEXT features achieves the best classification performance over the different programming languages.}

According to \tabref{combination_results}, NLP features alone 
achieve the lowest classification performance for both Java and Python, while we observe that this type of feature works well when dealing with Smalltalk class comments. 
On the one hand, class comments can often contain mixtures of structured (\eg code elements, such as class and attribute names) and unstructured information (\ie natural language). 
NEON
(i) leverages models trained on general-purpose natural language sentences to construct the parse tree of the sentences, and (ii) relies on the generated parse trees to identify common NLP patterns~\cite{Sorb19a}. 
Therefore, the presence of code elements degrades NEON's capability to generate accurate parse trees, and consequently complicates its pattern recognition task.     
On the other hand, code in Smalltalk resembles natural language (English) phrases, \eg the method, shown in \lstref{Code-snippets-MalTerms-class-smalltalk}, ``addString: using:'' takes two parameters ``string'' and ``CamelCaseScanner new'' written in a natural language sentence style. Similarly, class comments in Smalltalk are written in a more informal writing style (often using the first-person form, and written in complete sentences as shown in \figref{smalltalk-class-comment}), compared to Java and Python (which suggest to write in a formal way using the third-person form).
As demonstrated in previous work~\cite{Pani15a, Sorb15a}, the usage of predicate-argument patterns is particularly well-suited when dealing with classification problems in highly unstructured and informal contexts.

\vspace{2mm}
\begin{minipage}{\linewidth}
    \begin{lstlisting}[caption= {(Smalltalk) Code in MalTerms comment resembles natural language}, label={lst:Code-snippets-MalTerms-class-smalltalk}]
Terms subclasses Bag with support for handling stop words etc.

example: string
| terms |
terms := Terms new.
terms addString: string using: CamelCaseScanner new.
terms withCountDo: [ :term :count |term -> count ].
\end{lstlisting}
\end{minipage}

\finding{5}{When dealing with sentences containing mixtures of code elements and natural language texts, NLP tools based on parse trees fail to correctly identify well-suited NLP features. The usage of these features is otherwise recommended when the class comments are mostly unstructured.}
    
    \begin{table}[ht]
        \centering
        \small
        \captionsetup{font=small}
    \caption{Results for Java using the Random Forest classification model}
    \tablabel{best_model_category_results_Java}
    \begin{tabular}{l|lll}
        \hline
        \multirow{2}{*}{\textbf{Category}} & \multicolumn{3}{c}{NLP + TEXT} \\
        \cline{2-4}
         & Precision  & Recall & F-Measure \\
         \hline
         Summary      & 0.87       & 0.88   & 0.87 \\
         Expand       & 0.86       & 0.87   & 0.86 \\
         Ownership    & 0.99       & 0.99   & 0.99 \\
         Pointer      & 0.91       & 0.91   & 0.91 \\
         Usage        & 0.88       & 0.88   & 0.87 \\
         Deprecation  & 0.98       & 0.98   & 0.98 \\
         Rationale     & 0.95       & 0.95   & 0.95 \\
         \hline
    
    \end{tabular}
    \end{table}

For the sake of brevity, we base the following analysis on the NLP+TEXT configuration when used with the Random Forest classifier.
\tabref{best_model_category_results_Java},  \tabref{best_model_category_results_Python}, and \tabref{best_model_category_results_Pharo} respectively report the precision, recall, and F-Measure for the top frequent categories (see \secref{resultsRq2}) obtained through the Random Forest algorithm with the NLP+TEXT features.

In the case of Java, as shown in \tabref{best_model_category_results_Java}, \emph{Deprecation}, \emph{Ownership}, and \emph{Rationale} achieve high F-measure scores ( $\geq 95\%$), while \emph{Expand}, \emph{Summary} and \emph{Usage} are the categories with the lowest F-measure values (but still higher than 85\%). 
This means that for most Java categories Random Forest achieves very accurate classification results. 
However, we observe a certain variability in the results, depending on the category 
(see \tabref{all_projects_top_categories}).
While in our manually validated sample \emph{Summary} and \emph{Usage} occur more frequently than other categories, we observe that they achieve a classification performance lower than \emph{Deprecation} and \emph{Ownership} (\tabref{best_model_category_results_Java}).  
This outcome can be due to the presence of specific annotations or words that often occur in sentences belonging to the \emph{Deprecation} (\eg $@since$) and \emph{Ownership} (\eg $@author$) categories.
Although we removed all the special characters (annotation symbols) from the sentences, the techniques based on NLP+TEXT features can well capture the specific terms that frequently occur and are useful for identifying these categories.
For instance, in the Vaadin project, the \emph{Ownership} category always contains ``Vaadin'' in the $@author$ field.
Similarly, in the other Java projects, author name patterns are included in the NLP+TEXT feature set.

In contrast with \emph{Deprecation} and \emph{Ownership} categories, we do not observe recurrent annotations or words introducing sentences of the \emph{Rationale} type. 
However, they are more accurately classified compared to sentences in the \emph{Summary}, and \emph{Expand} categories. 
This could depend on the quantity and quality of the NLP features captured in these categories, jointly with a lower variability in the structure of comments falling in the \emph{Rationale} class.  
In particular, we observed that for the \emph{Rationale} category twelve unique NLP features have \emph{information gain} values higher than $0.01$, whereas two unique NLP features for the \emph{Summary} category and only one NLP feature for the \emph{Expand} category have \emph{information gain} scores higher than $0.01$.
In terms of quality, the top-ranked NLP feature (or pattern) of the \emph{Summary} category (\ie ``\emph{Represents [something]}'') occurs in only 3\% of the overall comments falling in this category,  whereas the top-ranked feature of the \emph{Rationale} category (\ie ``\emph{Has [something]}'') occurs in 8\% of the total comments belonging to such category.
Nevertheless, the NLP heuristics that occur in the sentences belonging to the \emph{Expand} class are also frequent in the instances of the \emph{Pointer} and \emph{Usage} categories, making it harder to correctly predict the type of the sentences falling in these categories.  
Specifically, the NLP feature with the highest \emph{information gain} score (\ie ``\emph{See [something]}'') for the \emph{Expand} category (information gain = 0.00676) is also relevant for identifying sentences of the \emph{Pointer} category, exhibiting an information gain value higher than the one observed for the \emph{Expand} category (\ie 0.04104).
    
\begin{table}[ht]
        \centering
        \small
        \captionsetup{font=small}
    \caption{Results for Python using the Random Forest classification model}
    \tablabel{best_model_category_results_Python}
    \begin{tabular}{l|ccc}
        \hline
        \multirow{2}{*}{\textbf{Category}} & \multicolumn{3}{c}{NLP + TEXT} \\
        \cline{2-4}
         & Precision  & Recall & F-Measure \\
         \hline
         Summary                            & 0.86       & 0.86   & 0.85      \\ 
         Usage                              & 0.83       & 0.83   & 0.82      \\
         Expand                             & 0.83       & 0.86   & 0.83      \\ 
         Development Notes                  & 0.87       & 0.89   & 0.87      \\
         Parameters                         & 0.86       & 0.86   & 0.85      \\
         \hline
    \end{tabular}
\end{table}

In the case of Python (see \tabref{best_model_category_results_Python}), 
the F-measure results are still positive ($>80\%$) for all the considered categories. 
Similar to Java, more frequent categories do not achieve the best performance.
For example, the category \emph{Parameter} is the least frequent among the categories considered but achieves higher performance than most of the categories, as shown in \tabref{best_model_category_results_Python}.  
In contrast to Java, Python developers frequently use specific words (\eg ``params'', ``args'', or ``note''), rather than annotations, to denote a specific information type.  We observe that these words frequently appear in sentences of the \emph{Parameter} and \emph{Development note} categories and these terms are captured in the related feature sets.
In the Python classification, the \emph{Usage} category reports the lowest F-measure due to its maximum ratio of incorrectly classified instances, \ie 17\% among all categories. 
This outcome can be partially explained by the small number of captured NLP heuristics (one heuristic \emph{``[something] defaults''} is selected according to the information gain measure with threshold 0.005).
We also observe that instances of the \emph{Usage} category often contain code snippets mixed with informal text, making it hard to identify features that would be good predictors of this class.
Similarly, the instances in the \emph{Expand} category also contain mixtures of natural language and code snippets.
Separating code snippets from natural language elements and treating each portion of the mixed text with a proper approach can help
(i) to build more representative feature sets for these types of class comment sentences, and 
(ii) to improve overall classification performance.

\begin{table}[ht]
    \centering
    \small
    \captionsetup{font=small}
\caption{Results for Smalltalk using the Random Forest classification model}
\tablabel{best_model_category_results_Pharo}
\begin{tabular}{l|ccc}
    \hline
    \multirow{2}{*}{\textbf{Category}} & \multicolumn{3}{c}{NLP + TEXT} \\
    \cline{2-4}
     & Precision  & Recall & F-Measure \\
     \hline
     Responsibility     & 0.79     & 0.82   & 0.78 \\
     Intent             & 0.92     & 0.92   & 0.90 \\
     Collaborator       & 0.83     & 0.94   & 0.83 \\
     Example            & 0.85     & 0.84   & 0.85 \\
     Class References   & 0.29     & 0.98   & 0.29 \\
    Key Messages       & 0.92     & 0.92   & 0.89 \\
    Key Implementation Points    & 0.87       & 0.89   & 0.85 \\
    \hline
\end{tabular}
\end{table}

Concerning Smalltalk, the Random Forest model provides slightly less stable results compared to Python and Java (see \tabref{combination_results}).
As shown in \tabref{best_model_category_results_Pharo}, in the case of Smalltalk,
\emph{Intent}
is the comment type with the highest F-measure. 
However, for most categories F-measure results are still positive ($>78\%$), except for the \emph{Class references} category. 
The \emph{Class references} category captures the other classes referred to in the class comment.
Random Forest is the ML algorithm that achieves the worst results (F-Measure of 29\%) for this category. However, the Naive Bayes algorithm achieves an F-Measure score of 93\% for the same category.
Similarly for the \emph{Collaborator} category the Naive Bayes model achieved better results compared to the Random Forest model. 
Both categories can contain similar information, \ie the name of other classes the class interacts with. 
We observe that in Smalltalk, camel case class names are generally split into separate words in comments, thus making it hard to identify them as classes from the text.
Nevertheless, as demonstrated in previous work~\cite{Pasc17a}, the Naive Bayes algorithm can achieve high performance in classifying information chunks in code comments containing code elements (\eg \emph{Pointer}), while its performance degrades when dealing with less structured texts (\eg \emph{Rationale}).
We observe similar behavior for the \emph{Links} category in Python taxonomy.
\figref{mapping-pharo-java-python} shows that all these categories such as  \emph{Links}, \emph{Pointer}, \emph{Collaborator}, and \emph{Class references} contain similar types of information.
In contrast, developers follow specific patterns in structuring sentences belonging to other categories such as \emph{Intent} and \emph{Example}, as reported in previous work~\cite{Rani21b}.
In future work, we plan to explore combining various machine learning algorithms to improve the results~\cite{Alex01a}.

To qualitatively corroborate the quantitative results and understand the importance of each considered NLP heuristic,
we computed the popular statistical measure
\emph{information gain} for the NLP features in each category, and ranked these features based on their scores. We use the default implementation of the information gain algorithm and the ranker available in Weka with the threshold value 0.005~\cite{Quin86a}.
Interestingly, for each category, the heuristics having the highest information gain values also exhibit easily explainable relations with the intent of the category itself. For instance, for the \emph{Responsibility} category in Smalltalk (which lists the responsibilities of the class) we observe that \emph{``Allows [something]''} and \emph{``Specifies [something]''} are among the best-ranked heuristics. Similarly, we report that \emph{``[something] is used''} and \emph{``[something] is applied''} are among the heuristics having the best information gain values for the \emph{Collaborator} category (\ie interactions of the class with other classes), while the heuristics \emph{``[something] is class for [something]''} and \emph{``Represents [something]''} have higher information gain values when used to identify comments of the \emph{Intent} type (\ie describing the purpose of the class).
These heuristics confirm the patterns identified by Rani \etal in their manual analysis of Smalltalk class comments~\cite{Rani21b}.
Similar results are observed for the other languages.
More specifically, for the \emph{Summary} category in Java, the heuristics \emph{``Represents [something]''} and \emph{``[something] tests''} are among the NLP features with the highest information gain.
Instead, in the case of the \emph{Expand} and \emph{Pointer} categories, we observe a common relevant heuristic: \emph{``See [something].''}.
The analysis of the top heuristics for the considered categories highlight that developers follow similar patterns (\eg \emph{``[verb] [something]''}) to summarize the purpose and responsibilities of the class across the different programming languages. However, no common patterns are found when discussing specific implementation details (\emph{Expand} and \emph{Usage} in Java and Python and \emph{Example} in Smalltalk).

\finding{6}{In all the considered programming languages, developers follow similar patterns to summarize the purpose and the responsibilities of a class. No common patterns are observed when implementation details are discussed.}


\textbf{\emph{Discussion:}} To further confirm the reliability of our results, we complement previous results with relevant statistical tests. 
In particular, the \emph{Friedman test} reveals that the differences in performance among the classifiers is statistically significant in terms of F-Measure. 
Thus, we can conclude that when comparing the performance of classifiers and using different input configurations, the choice of the classifier significantly affects the results. 
Specifically, to gain further insights about the groups that statistically differ, we performed the Nemenyi test. 
Results of the test suggest that the Naive Bayes and the J48 models do not statistically differ in terms of F-Measure, while the Random Forest model is the best performing model, with statistical evidence ($p-value < 0.05$).

To analyze how the usage of different features (NLP, TEXT, NLP+TEXT) affects the classification results, we also executed a Friedman test on the F-Measure scores obtained by the Random Forest algorithm for each possible input combination. 
The test concluded that the difference in the results with different inputs is statistically significant ($p-value < 0.05$). 
To gain further insight into the groups that statistically differ, we performed a Nemenyi test. 
The test revealed that there is significant difference between the NLP and NLP+TEXT combinations ($p-value < 0.05$). 
This result confirms the importance of both NLP and TEXT features when classifying class comment types in different languages.
The input data and the scripts used for the tests are provided in the Replication package.\footnote{\repFolder{Result/RQ2/Statistical-analysis}}
Currently, we use the tf-idf weighting scheme for TEXT features but we plan to experiment with other weighting schemes for future work, for instance, TF-IDF-ICSDF (Inverse Class Space Density Frequency) as it considers also the distribution of inter-class documents when calculating the weight of each term\cite{Doga20a}.

\subsection{Reproducibility}
In order to automate the study, we developed a Command Line Interface (CLI) application in Java.
The application integrates the external tool NEON, and the required external libraries (Standford NLP) to process and analyze the data (WEKA) using Maven.
The input parameters for the application are the languages (\eg Java, Python) to analyze, their input dataset path, and the tasks to perform.
Various tasks fetch the required input data from the database, perform the analysis, and store the processed output data back in it.
The results of intermediate steps are stored in the database\footnote{\repFile{Results/RQ2/All-steps-result.sqlite}} and the final results are exported as CSV automatically using \emph{Apache Commons CSV}.

\section{Threats to validity}
\seclabel{Threats-to-validity}
We now outline potential threats to the validity of our study.
\emph{Threats to construct validity}
mainly concern the measurements used in the evaluation.
To answer the RQs, we did not consider the full ecosystem of projects in each language but selected a sample of projects for each language. 
To alleviate this concern to some extent, we selected heterogeneous projects used in the earlier comment analysis work of Java, Python, and Smalltalk~\cite{Pasc17a,Zhan18a,Rani21b}.
The projects in each language focus on different domains such as visualization, data analysis, or development frameworks.
They originate from different ecosystems, such as Google and Apache in Java, or Django Foundation, or community project in Python.
Thus, the projects follow different comment guidelines (or coding style guidelines). 
Additionally, the projects are developed by many contributors (developers), which further lowers the risk towards a specific developer commenting style.

Another important issue could be due to the fact that we sampled only a subset of the extracted class comments.
However, the sample size limits the estimation imprecision to 5\% of error for a confidence level of
95\%. 
To further mitigate concerns related to subjectiveness and bias in the evaluation, the truth set was built based on the judgement of four annotators  (four authors of this work) who manually analyzed the resulting sample.
Moreover, an initial set of 50 elements for each language was preliminarily labelled by all annotators and all disagreements were discussed between them.
To reduce the {likelihood} that the chosen sample comments are not representative of the whole population, we used a stratified sampling approach to choose the sample comments from the dataset, thus considering the quartiles of the comment distribution for each language.\footnote{\repFile{Dataset/RQ1/Java/projects-distribution.pdf}}

Another threat to construct validity concerns the definition of the \emph{CCTM} and the mappings of the different language taxonomies performed by four human subjects. To counteract this issue, we used the categories defined by the earlier works in the comment analysis~\cite{Pasc17a,Zhan18a,Rani21b}.

\emph{Threats to internal validity} concern confounding factors that could influence our results.
The main threat to internal validity in our study is related to the manual analysis carried out to prepare the \emph{CCTM} and the mapping taxonomy. {Since it is} performed by human subjects, it could be biased.  
Indeed, there is a level of subjectivity in deciding whether a comment type belongs to a specific category of the taxonomy or not, and whether a category of one language taxonomy maps to a category in another language taxonomy or not.
To counteract this issue, the evaluators of this work were two Ph.D.\ candidates and two faculty members, each having at least four years of programming experience.
All the decisions made during the evaluation process and validation steps are reported in the replication package (to provide evidence of the non-biased evaluation), and described in detail in the paper.\footnote{\repFolder{Result/RQ1/Manually-classified-comments}}
Also, we performed a two-level validation step.
This validation step involved further discussion among the evaluators, whenever their opinions diverged, until they reached a final consensus.

{\emph{Threats to conclusion validity} concern the relationship between treatment and outcome. Appropriate statistical procedures have been adopted to draw our conclusions.
{To answer $RQ_2$, we investigate whether the differences in the performance achieved by the different machine learning models with different combination of features were statistically significant.  
To perform this task, we used the Friedman test, followed by a post-hoc Nemenyi test, as recommended by Dem{\v{s}}ar~\cite{Dems06a}. }

\emph{Threats to  external validity} concern the generalization of our results.
The main aim of this paper is to investigate the class commenting practices for the selected programming languages.
The proposed approach may achieve different results in other programming languages or projects.
To limit this threat, we considered both static and dynamic types of object-oriented programming languages having different commenting style and guidelines.  
To reduce the threat related to the project selection, we chose diverse projects, used in previous studies about comment analysis and assessment.
The projects vary in terms of size, domain, contributors, and ecosystems.
Finally, during the definition of our taxonomy (\ie  \emph{CCTM}) we mainly rely on a quantitative analysis of class comments, without directly involving the actual developers of each programming language.
Specifically, for future work, we plan to involve developers, via surveys and interviews.
This step is particularly important to improve the results of our work and to design and evaluate further automated approaches that can help developers achieve a high quality of comments.

\section{Related Work}
\seclabel{Related-work}

\textbf{Comment Classification}. 
Comments contain various information types~\cite{Ying05a} useful to support numerous software development tasks.
Recent work has categorized the links found in comments~\cite{Hata19a}, and proposed comment categories based on the actual meaning of comments~\cite{Padi09a}.
Similar to these studies, our work is aimed at supporting  developers in discovering important types of information from class comments. 

Steidl \etal classified the comments in Java and C/C++ programs automatically using machine learning approaches.
The proposed categories are based on the position and syntax of the comments, \eg inline comments, block comments, header comments \etc~\cite{Stei13b}.
Differently from Steidl \etal, our work focuses on analyzing and identifying semantic information found in class comments  in Java, Python, and Smalltalk.
Pascarella \etal presented a taxonomy of code comments for Java projects~\cite{Pasc17a}.
In the case of Java, we used the taxonomy from Pascarella \etal to build our Java CCTM categories. 
However, our work is different from the one of Pascarella \etal, as it focuses on class comments in three different languages, which makes our work broader in terms of studied languages and more specific in terms of type of code comments studied.
Complementary to Pascarella \etal's work, Zhang \etal~\cite{Zhan18a} reported a code comment taxonomy in Python. 
Compared to the Python comment taxonomy of Zhang \etal, we rarely observed the \emph{Version}, \emph{Todo}, and \emph{Noise} categories in our Python class comment taxonomy.
More importantly, we found other types of information in Python class comments that developers {embed} in the class comments but were not included in the Python comment taxonomy of Zhang \etal such as the \emph{Warning}, \emph{Observation}, \emph{Recommendation} and \emph{Precondition} categories of the proposed CCTM.
More in general, our work complement and extend the  studies of Pascarella \etal and Zhang \etal by focusing on class comments in three different languages, which makes our work broader in terms of studied languages as well as the types of code comments reported and automatically classified. 

Several studies have experimented with numerous approaches to identify the different types of information in comments~\cite{Drag10a,Ying14a,Shin18a,Geis20a}.
For instance, Dragan \etal used a rule-based approach to identify the Stereotype of a class based on the class signature~\cite{Drag10a}.
Their work is aimed at recognizing the class type (\eg data class, controller class) rather than the type of information available within class comments, which is the focus of our work.
Shinyama \etal ~\cite{Shin18a} focused on discovering specific types of local comments (\ie  explanatory comments) that explain how the code works at a microscopic level inside the functions.
Similar to our work, Shinyama \etal and Geist \etal considered recurrent patterns, but crafted them manually as extra features to train the classifier.
Thus, our approach is different, as it is able to \emph{automatically} extract  different natural language patterns (heuristics), combining them with other textual features, to classify class comment types of different languages.


\textbf{Further Comment Analysis}. 
Apart from identifying information types within comments, analyzing comments for other purposes has also gained a lot of attention in the research community in the past years. 
Researchers are investigating methods and techniques for generating comments~\cite{Haid10a,Niel19a}, assessing their quality~\cite{Kham10a,Stei13b,Yu16a}, detecting inconsistency between code and comments~\cite{Rato17a,Wen19a,Zhou17a,Liu18a}, examining co-evolution of code and comments~\cite{Jian06a,Flur07b,Flur09a,Ibra12a}, identifying bugs using comments~\cite{Tan07b}, or establishing traceability between code and comments~\cite{Marc03b,Anto00c}.

Other work has focused on experimenting with rule-based and machine-learning-based techniques and identified further program comprehension challenges. 
For instance, for labeling source code, De Lucia \etal~\cite{DeLu12a} found that simple heuristic approaches work better than more complex approaches, \eg LSI and LDA. 
Moreno \etal~\cite{More13c} proposed NLP and heuristic-based techniques  to generate summaries for Java classes. 
Finally, recent research by Fucci \etal~\cite{Fucc19a} studied how well modern text classification approaches can identify the information types in API documentation automatically. 
Their results have shown how neural network outperforms traditional machine learning approaches and naive baselines in identifying multiple information types (multi-label classification). 
In the future, we plan to experiment with class comment classification using neural networks.

\section{Conclusion}
\seclabel{conclusion}
Class comments provide a high-level understanding of the program and help one to understand a complex program.
Different programming languages have their own commenting guidelines and notations, thus identifying a particular type of information from them for a specific task or evaluating them from the content perspective is a non-trivial task.

To handle these challenges, we investigate class commenting practices of a total of 20 projects from three programming languages, Java, Python, and Smalltalk.
We identify more than 17 types of information class comments of each programming language.
To automatically identify the most frequent information types characterizing these languages, we propose an approach based on natural language processing and text analysis that classifies, with high accuracy, the most frequent information types of all the investigated languages.

The contributions of our work are
(i) an empirically validated taxonomy for class comments in three programming languages;
(ii) a mapping of taxonomies from previous works;
(iii) a common automated classification approach able to classify class comments according to a broad class comment taxonomy using various machine learning models trained on top of different feature sets; and
(iv) a publicly available dataset of 37\,446 class comments from 20 projects and 1\,066 manually classified class comments\footnote{\repFolder{Dataset/RQ1/Java}}.

Our results highlight the different kinds of information class comments contain across languages.
We found many instances of specific information types that are not suggested or mentioned by their respective coding style guidelines.
To what extent developer commenting practices adhere to these guidelines is not known yet and is part of our future work agenda.
We argue that such an analysis can help in evaluating the quality of comments, also suggested by previous works~\cite{Padi09a,Haou11a,Stei13b}.
To investigate this aspect, we plan to extract the coding style guidelines of heterogeneous projects related to comments and compare the extracted guidelines with the identified comment information types (using our proposed approach).
Moreno \etal used a template-based approach to generate Java comments~\cite{More17a} where the template includes specific types of information they deem important for developers to understand a class.
Our results show that frequent information types vary across systems.
Using our approach to identify frequent information types, researchers can customize the summarization templates based on their software system.

\section*{Acknowledgments}
We gratefully acknowledge the financial support of the Swiss National Science Foundation for the project
``Agile Software Assistance'' (SNSF project No.\ 200020-181973, Feb.\ 1, 2019 - April 30, 2022).
We also thank Ivan Kravchenko for helping us to extract the data for Java and Python.

\bibliographystyle{elsarticle-num}
\interlinepenalty=10000 
\bibliography{scg}

\end{document}